\documentclass{article}
\usepackage{epsfig}

\date{\today}

\usepackage[intlimits,tbtags]{amsmath} 
\usepackage{amsfonts,amssymb}
\usepackage{amsthm}
\usepackage{amscd}

\usepackage{amsmath}%
\usepackage{amsfonts}%
\usepackage{amssymb}%
\usepackage{graphicx}
\usepackage{subfig}
\usepackage{anysize}
\usepackage{inputenc}
\usepackage{longtable}
\usepackage{hyperref}
\usepackage[toc,page]{appendix}
\usepackage{booktabs}
\usepackage{url}
\usepackage[nottoc]{tocbibind}
\usepackage{xcolor}

\begin{document}
\title{Stability of superconducting strings coupled to cosmic strings}
\author{{\large Alexandru Babeanu \footnote{email: ababeanu@jacobs-alumni.de} } and 
{\large Betti Hartmann \footnote{email: b.hartmann@jacobs-university.de}}
\\ \\
{\small School of Engineering and Science, Jacobs University Bremen, 28759 Bremen, Germany}  }

\date{}
\newcommand{\dd}{\mbox{d}}
\newcommand{\tr}{\mbox{tr}}
\newcommand{\la}{\lambda}
\newcommand{\ka}{\kappa}
\newcommand{\f}{\phi}
\newcommand{\vf}{\varphi}
\newcommand{\F}{\Phi}
\newcommand{\al}{\alpha}
\newcommand{\ga}{\gamma}
\newcommand{\de}{\delta}
\newcommand{\si}{\sigma}
\newcommand{\bomega}{\mbox{\boldmath $\omega$}}
\newcommand{\bsi}{\mbox{\boldmath $\sigma$}}
\newcommand{\bchi}{\mbox{\boldmath $\chi$}}
\newcommand{\bal}{\mbox{\boldmath $\alpha$}}
\newcommand{\bpsi}{\mbox{\boldmath $\psi$}}
\newcommand{\brho}{\mbox{\boldmath $\varrho$}}
\newcommand{\beps}{\mbox{\boldmath $\varepsilon$}}
\newcommand{\bxi}{\mbox{\boldmath $\xi$}}
\newcommand{\bbeta}{\mbox{\boldmath $\beta$}}
\newcommand{\ee}{\end{equation}}
\newcommand{\eea}{\end{eqnarray}}
\newcommand{\be}{\begin{equation}}
\newcommand{\bea}{\begin{eqnarray}}

\newcommand{\ii}{\mbox{i}}
\newcommand{\e}{\mbox{e}}
\newcommand{\pa}{\partial}
\newcommand{\Om}{\Omega}
\newcommand{\vep}{\varepsilon}
\newcommand{\bfph}{{\bf \phi}}
\newcommand{\lm}{\lambda}
\def\theequation{\arabic{equation}}
\renewcommand{\thefootnote}{\fnsymbol{footnote}}
\newcommand{\re}[1]{(\ref{#1})}
\newcommand{\R}{{\rm I \hspace{-0.52ex} R}}
\newcommand{\N}{{\sf N\hspace*{-1.0ex}\rule{0.15ex}%
{1.3ex}\hspace*{1.0ex}}}
\newcommand{\Q}{{\sf Q\hspace*{-1.1ex}\rule{0.15ex}%
{1.5ex}\hspace*{1.1ex}}}
\newcommand{\C}{{\sf C\hspace*{-0.9ex}\rule{0.15ex}%
{1.3ex}\hspace*{0.9ex}}}
\newcommand{\eins}{1\hspace{-0.56ex}{\rm I}}
\renewcommand{\thefootnote}{\arabic{footnote}}

\maketitle


\bigskip

\begin{abstract}
We study the stability of superconducting strings in a 
U(1)$_{\rm local}$ $\times$U(1)$_{\rm global}$ model coupled
via a gauge field interaction term to U(1) Abelian-Higgs strings.  
The effect of the interaction on current stability is numerically 
investigated by varying the relevant 
parameters within the physical limits of our model. We find that the
propagation speed of transverse (resp. longitudinal) perturbations
increases (decreases) with increasing binding between the superconducting and
Abelian-Higgs string. Moreover, we observe that for small enough width of the flux tube of the
superconducting string and/or large enough interaction between the superconducting
and the Abelian-Higgs string superconducting strings cannot carry space-like, 
i.e. magnetic currents. Our model can be seen as a field theoretical realization
of bound states of p F-strings and q superconducting D-strings and has important implications
to vorton formation during the evolution of networks of such strings.

\end{abstract}
\medskip
\medskip
 \section{Introduction}
 \label{s1}

Cosmic strings are very massive topological defects which could have formed via the Kibble mechanism \cite{kibble} 
during one of the phase transitions in the early universe 
such as the Grand  Unification phase transition or the electroweak phase transition. 
They are analogous to defects in condensed matter physics, 
such as flux tubes in superconductors or vortex filaments in superfluid helium \cite{CS} 
being thus 
intimately connected to spontaneous symmetry breaking. The study of cosmic strings is currently motivated 
by the possibility that their production is related to inflation models resulting from 
String theory \cite{polchinski}.
Brane inflation is a popular inflationary model that
can be embedded into String Theory and 
predicts the formation of cosmic string networks at the end of inflation \cite{braneinflation}.
E.g. in the framework of type IIB String Theory the inflaton field corresponds to the
distance between two Dirichlet branes with 3 spatial dimensions (D$3$-branes) and inflation ends
when these two branes collide and annihilate. The production of strings (and lower dimensional branes)
then results from the collision of these two branes.
Each of the original D3-branes has a U(1) gauge symmetry that gets broken when
the branes annihilate.
If the gauge combination is Higgsed, magnetic
flux tubes of this gauge field carrying Ramond-Ramond (R-R) charge are D-branes with one spatial dimension, 
so-called
D-strings. When the gauge combination is confined the field is condensated into 
electric flux tubes carrying Neveu Schwarz-Neveu Schwarz (NS-NS) charges and these objects are fundamental strings (F-strings) 
\cite{dvali_vilenkin}. 
D-strings and F-strings are so-called cosmic superstrings \cite{polchinski} 
which seem to be a generic prediction of supersymmetric 
hybrid inflation \cite{lyth} and grand unified based inflationary models \cite{jeannerot}. 
D- and F-strings, however, have different properties than the usual (solitonic) cosmic strings. 
The probability of intercommutation 
of solitonic strings is equal to one but less than one in the case of cosmic superstrings.
Therefore, solitonic strings do not merge, while cosmic superstrings tend to form bound states. 
When p F-strings and 
q D-strings interact, they can merge and 
form bound states, so-called (p,q)-strings \cite{copeland_myers_polchinski} whose
properties have been investigated \cite{bulk}. 
Even though the origin of (p,q)-strings is type IIB string theory, 
their properties can be investigated in the framework of field theoretical models
\cite{saffin,rajantie,salmi,urrestilla}. 
The influence of gravity on field theoretical (p,q)-strings has been studied in 
\cite{hartmann_urrestilla}. 

The field theoretical model most frequently used to study cosmic strings is the U(1) Abelian-Higgs model \cite{NO}, which possesses
static, line-like solutions. These solutions are straight strings in the sense that their fields do not depend
on the $z$-coordinate. These straight strings can be interpreted as a local description for potentially curved objects, 
when viewed at large scales. In fact, it is believed that they are part of intricate, dynamical networks of 
interacting cosmic strings \cite{CSN}, which may bend, collide and eventually form a multitude of loops. 
If strings have no internal structure, as in the U(1) Abelian-Higgs model, these cosmic string loops would rapidly 
disappear through self-gravitational collapse, radiating all their energy away \cite{CSLC}.  
Nonetheless, string loops may survive if they carry a current in their core. Such current-carrying string loops 
are called vortons and, in the simplest scenario, 
they reach a stable, ring-like shape \cite{VF0,VF0a,VF0b}. Further studies 
showed that vortons could have been produced from cosmic strings appearing 
during the electroweak phase transition \cite{VF1,VF2}. In order to understand the
cosmological implications of such objects it is essential to know about their stability
and is has been suggested that vortons can be unstable under certain conditions \cite{instability_vortons}.
A current carrying string can be most simply described using a U(1)$\times$U(1) model \cite{Witten}. In this case 
the U(1) Abelian-Higgs model
is coupled to a U(1) Abelian-scalar field model through a potential term. The U(1) symmetry of the latter is unbroken, which leads
to a locally conserved Noether current and a globally conserved Noether charge. If the U(1) symmetry is gauged as has been
the case in the original proposal \cite{Witten}, the energy 
density per unit length diverges due to the logarithmic fall-off
of the electromagnetic field. It has hence been suggested to consider 
the unbroken U(1) symmetry to be global or equivalently the
corresponding gauge coupling to be small in order to have localized superconducting strings \cite{PP}.
In this limit it is possible to apply the formalism developed in 
\cite{carter1,carter2,carter3}. This distinguishes between superconducting strings 
with time-like currents (``electric'' regime)
and space-like currents (``magnetic'' regime) and relates 
the longitudinal and transversal perturbations on the string
to the energy and tension per unit length giving explicit criteria for 
the stability of these objects. In \cite{PP} it was shown
that within the full U(1)$\times$U(1) model longitudinal perturbations 
always propagate more slowly than transverse ones.
Moreover, a general logarithmic equation of state was suggested in \cite{PP} 
which has been confirmed numerically
to hold \cite{BHBC}.  

Recent studies have explored a possibility of coupling two U(1) Abelian-Higgs models (with both U(1) symmetries broken)
via a gauge field interaction term \cite{BHFA}. This interaction
has been motivated by the possible existence of a separate, dark matter sector in the universe \cite{dark_model}, which allows 
for the existence of so called ``dark strings'' \cite{DS}. 
This dark sector would weakly interact with the Standard Model sector 
through a small gauge field interaction and is motivated by recent astrophysical observations \cite{observation_ee} that
have shown an excess electronic production in the galaxy with electrons having energies between a few GeV and a few TeV. 
In \cite{BHFA}, the interaction of dark strings with U(1) Abelian-Higgs strings was 
studied and the possibility of forming bound states was explored. This study was extended to semilocal strings \cite{BHYB} which are
solutions of a SU(2)$\times$U(1) model with an ungauged SU(2) symmetry.

In this paper we study bound states of a U(1)$_{\rm local}$ $\times$U(1)$_{\rm global}$ 
superconducting string and an Abelian-Higgs string.
We will put the emphasize on the study of the stability of the superconducting strings 
when these are coupled
via an attractive gauge interaction to Abelian-Higgs strings. Our model can be
seen as a toy model for (p,q)-strings in which the attractive interaction mediated
by the scalar fields as suggested in the original model \cite{saffin} is replaced
by an attractive interaction mediated by the gauge fields. D-strings can
carry currents. In \cite{davisetal} the formation of vortons from superconducting
D-strings carrying chiral fermion zero modes has been discussed, while it
has been suggested that D-strings could also carry bosonic currents in special cases, e.g.
if a D-string would be coincident with a D7-brane \cite{polchinski}. In the following we
will hence assume that the superconducting string in the bound state is a superconducting
D-string. 

Our paper is organized as follows: we give the model in section \ref{model}. We present our results in section \ref{results}
and conclude in section \ref{conclusions}.

\section{The model}
\label{model}

From a qualitative perspective the field theoretical model studied in this paper 
is a combination of the models
studied in \cite{PP,BHBC,BHFA}. For the vanishing-current limit it can also be seen
as an alternative to the field theoretical toy model for (p,q)-strings suggested in \cite{saffin}. In the
latter case, the attractive interaction between the p- and the q-string is mediated
via a scalar field potential term, while here the attractive interaction is mediated
via a gauge field term. The model without currents has been studied in \cite{BHFA}. 
Here, we extend this investigation assuming that one of the strings possesses a 
current. Our model could hence be a toy model for (p,q)-strings where the q D-strings
are superconducting.  

The field theoretical model has the following action
\begin{equation}
\label{action}
 S=\int d^4 x \sqrt{-g} {\mathcal L} \ ,
\end{equation}
where $g$ denotes the determinant of the metric with signature $(+---)$ and $\mathcal{L}$ is the Lagrangian density given by
\begin{equation}
	\label{LagrGen}
	\begin{split}
	\mathcal{L} =	
			&	- \frac{1}{4} F^{(1)}_{\mu\nu} F_{(1)}^{\mu\nu} 
				- \frac{1}{4} F^{(2)}_{\mu\nu} F_{(2)}^{\mu\nu} 
		         + \frac{\varepsilon_2}{2} F^{(1)}_{\mu\nu} F_{(2)}^{\mu\nu}
		 \\
			&	+ (D_{\mu}\phi_{1})^{\dagger} (D^{\mu}\phi_{1})
				+ (D_{\mu}\phi_{2})^{\dagger} (D^{\mu}\phi_{2})
				+ (\partial_{\mu}\phi_{3})^{\dagger} (\partial^{\mu}\phi_{3}) \\
			&	- \frac{\lambda_1}{4}(|\phi_1|^2 - \eta_1^2)^2
				- \frac{\lambda_2}{4}(|\phi_2|^2 - \eta_2^2)^2
				- \frac{\lambda_3}{4}(|\phi_3|^2 - 2\eta_3^2) |\phi_3|^2
				- \lambda_{4}|\phi_2|^2 |\phi_3|^2 \ , \\     
	\end{split}
\end{equation}
where $F^{(a)}_{\mu\nu}=\partial_{\mu} A^{(a)}_{\nu}- \partial_{\nu} A^{(a)}_{\mu}$ 
is the field strength
tensor of the U(1) gauge field $A^{(a)}_{\mu}$, $a=1,2$ and 
$D_{\mu} \phi_{a} =\partial_{\mu} \phi_{a} - i e_a A^{(a)}_{\mu} 
\phi_{a}$ is the
covariant derivative of the complex scalar field $\phi_{a}$, $a=1,2$. The current carrying field
$\phi_3$ is ungauged in order to have localized current-carrying strings. 
$e_a$, $a=1,2$ and $\lambda_i$, $\eta_i$, $i=1,2,3$ denote
the gauge couplings, the self-couplings and the vacuum expectation values, respectively.
The index (1) refers to the fields and coupling constants associated to the 
U(1) Abelian-Higgs model describing e.g. the p F-strings in a (p,q)-string. The indices
(2) and (3) are related to the fields and coupling constants of the 
U(1)$_{\rm local}\times$U(1)$_{\rm global}$ model describing superconducting strings, e.g.
the q D-strings in a (p,q)-string.
In the following (see
Ansatz below) the index (2) will denote the fields of the broken U(1) symmetry, 
while the index (3) will denote the unbroken
symmetry of the superconducting string. $\varepsilon_2$ is the gauge interaction coupling 
between the U(1) Abelian-Higgs string and the broken U(1) symmetry of the superconducting string.
It can only take values $\varepsilon_2 \in [0.0,1.0)$ for mathematical reasons 
and $\varepsilon_2 \in [0.0,0.001]$ 
for experimental reasons described by \cite{BHFA} in the context of dark strings.
The interactions between the two scalar fields of the superconducting string, 
which is responsible for the existence of the 
current, is here mediated by the constant $\lambda_4$. 
Note that the constant $(\lambda_3 \eta_3^4)/4$ in the quartic potential 
of $|\phi_3|$ has been eliminated from the model, 
as it would simply shift the energies by a constant term.
The Higgs and gauge field masses are $m_{\rm H,i}=\sqrt{\lambda_i} \eta_i$ 
and $m_{\rm W,i}=e_i \eta_i$, $i=1,2$, respectively.
The mass of the scalar field $\phi_3$ associated to 
the unbroken U(1) is given by $m_{\rm S}=\sqrt{2\lambda_4 \eta_2^2 - \lambda_3 \eta_3^2}$. 
The two strings each possess a scalar core with width $\rho_{\rm H,i}\propto m_{\rm S,i}^{-1}$ and
a flux/gauge tube with width $\rho_{\rm W,i}\propto m_{\rm W,i}^{-1}$.

We use the following cylindrically symmetric ansatz for the scalar fields
\begin{align}
                \label{ScalAn}
       & \phi_{1}(\rho,\varphi) = \eta_1 f_1(\rho) e^{in_1\varphi}  \ \ , \ \ 
\phi_{2}(\rho,\varphi) = \eta_2 f_2(\rho) e^{in_2\varphi} \ \ , \ \ 
\phi_{3}(\rho,\varphi) = \eta_3 f_3(\rho) e^{i(\omega t + kz)} \nonumber 
\end{align}
and the following ansatz for the associated gauge fields
\begin{equation}
                \label{GaugeAn}
        A^{(a)}_{\mu}(\rho)dx^{\mu} = \frac{1}{e_a}(n_a - P_a(\rho)) d\varphi 
\end{equation}
with $a=1,2$. Here the $n_a\in \mathbb{Z}$, $a=1,2$ are integers indexing the vorticity 
of the fields. Note that if we think of our model as a toy model for a (p,q)-string
with p F-strings and q superconducting D-strings, then $n_1\equiv$ p and $n_2\equiv$ q.
The strings possess magnetic fields $\vec{B}_a=B_a\vec{e}_z$ with $B_a=-P'_a/(e_a \rho)$ 
and magnetic fluxes $\Phi_a=2\pi n_a/e_a$.

The 4-current can be computed from Noether's theorem \cite{NMPS}, 
which states that if under an infinitesimal transformation $\phi \rightarrow \phi + \alpha\delta \phi$, where $\alpha$ is a
parameter
the Lagrangian density is invariant up to surface terms, i.e.
\begin{equation}
\mathcal{L} \rightarrow \mathcal{L} + \alpha \partial_{\mu} K^{\mu}
\end{equation}
then there exists a locally conserved 4-current given by
\begin{equation}
\label{NT}
J^{\mu} = \frac{\partial \mathcal{L}}{\partial(\partial_{\mu}\phi)} \delta \phi - K^{\mu}  \ .
\end{equation}
The explicit expressions in our model read
\begin{equation}
	\label{NC}
 J^t = \omega q_3^2 f_3^2 \ \ \ , \ \ \  J^z = -k q_3^2 f_3^2  \ ,
\end{equation}
where the integral of $J^t$ and $J^z$ over $x$ and $\varphi$ give the Noether charge and total Noether current per unit length, 
respectively, which are globally conserved. It can be easily seen that the value of the condensate $f_3(0)$ controls the amount of 
both charge and current per unit length, while the quadratic norm $\chi^2$ determines the nature of the 4-current. 
According to the criterion in \cite{carter1,carter2,carter3}, we will call the 4-current time-like (electric) when $\chi^2 < 0$
since $k^2 < \omega^2$, space-like (magnetic) when $\chi^2 > 0$ since $k^2 > \omega^2$ or light-like when $\chi^2 = 0$ 
since $k^2 = \omega^2$. 
This terminology is related to the fact that $k^2$ ($\omega^2$)
vanishes when convenient Lorentz boosts are applied in the time-like (space-like) case. 
Note also
that the light-like case includes $\omega=k=0$, but is not restricted to it.

The energy-momentum tensor 
$T_{\nu}^{\mu}=2g^{\mu\sigma} 
\frac{\partial {\cal L}}{\partial g^{\sigma\nu}} - \delta_{\nu}^{\mu} {\cal L}$
is given by
\begin{equation}
 T_{\nu}^{\mu}=-\sum_{i=1}^2 F^{\mu\sigma}_{(i)} F_{\nu\sigma}^{(i)} + 2\varepsilon_2 F^{\mu\sigma}_{(1)} F_{\nu\sigma}^{(2)}
+ 
\sum_{i=1}^2 \left[(D^{\mu}\phi_i)^{\dagger} D_{\nu} \phi_i + 
(D_{\nu}\phi_i)^{\dagger} D^{\mu} \phi_i\right] + (\partial_{\nu}\phi_3)^{\dagger} 
\partial^{\mu} \phi_3 -\delta^{\mu}_{\nu} {\cal L}  \ .
\end{equation}

It is convenient to use the following rescaled quantities: 
\begin{align}
\label{resc}
       & x=e_1\eta_1\rho  \ \ , \ \ \beta_i = \frac{\lambda_i}{e_1^2} \ , \ i=1,2,3,4 \  \ ,  \ \ \chi^2 = \frac{k^2-\omega^2}{e_1^2\eta_1^2}  \ , \\
       & g_2 =\frac{e_2}{e_1}  \ \ ,  \ \ q_2 =\frac{\eta_2}{\eta_1} \ \ , \ \ q_3 =\frac{\eta_3}{\eta_1} \ .
\end{align}
Then, rescaling the Lagrangian density as $\mathcal{L}\rightarrow \mathcal{L}/(e_1^2 \eta_1^4)$ it reads
\begin{equation}
	\label{LagrPartic2}
		\begin{split}
\mathcal{L} = 
     &	-\frac{1}{2x^2}\left[ P_1'^2 + \frac{P_2'^2}{g_2^2} \right] + \frac{\varepsilon_2}{g_2} \frac{P_1' P_2'}{x^2}  \\
     & 	-\left[f_1' + \frac{P_1^2 f_1^2}{x^2}\right]
	-q_2^2\left[ f_2' + \frac{P_2^2 f_2^2}{x^2}\right] 
     	-q_3^2\left[\chi^2 f_3^2 + f_3' \right] \\
     &	-\frac{\beta_1}{4} \left(f_1^2-1\right)^2
	-q_2^4\frac{\beta_2}{4} \left(f_2^2-1\right)^2
	-q_3^4\frac{\beta_3}{4}  \left(f_3^2 - 2\right)f_3^2 \\
     &	-q_2^2 q_3^2\beta_4  f_2^2 f_3^2 \ , \\
		\end{split}
\end{equation}
where the dependence of all functions on the rescaled radial distance $x$ has been omitted for simplicity and the prime
now and in the following denotes the derivative with respect to $x$.
In these rescaled variables the energy $E$ and tension $T$ per unit length of these solutions are given by
\begin{equation}
	\label{ETdens}
		E = 2\pi \displaystyle \int_0^{\infty} T_t^t x d x  \ \ \ \ , \ \ \ \ 
		T = 2\pi \displaystyle \int_0^{\infty} T_z^z x d x  \ ,
\end{equation}
where the energy density $T_t^t$ and tension density $T_z^z$ are the $tt$- and $zz$-components of the
energy momentum tensor $T_{\mu}^{\nu}$ 
 and read
\begin{equation}
	\label{EnergPartic}
		\begin{split}
T_t^t = 
     &	\frac{1}{2x^2}\left[ P_1'^2 + \frac{P_2'^2}{g_2^2} \right] - \frac{\varepsilon_2}{g_2} \frac{P_1' P_2'}{x^2} \\
     & 	+\left[f_1' + \frac{P_1^2 f_1^2}{x^2}\right]
	+q_2^2\left[ f_2' + \frac{P_2^2 f_2^2}{x^2}\right] 
     	+q_3^2\left[\bar{\chi}^2 f_3^2 +  f_3'\right] \\
     &	+\frac{\beta_1}{4}  \left(f_1^2-1\right)^2
	+q_2^4\frac{\beta_2}{4}  \left(f_2^2-1\right)^2
	+q_3^4\frac{\beta_3}{4}  \left(f_3^2 - 2\right)f_3^2 \\
     &	+q_2^2 q_3^2\beta_4  f_2^2 f_3^2 \\
		\end{split}  
\end{equation}
and
\begin{equation}
	\label{TensPartic}
		\begin{split}
T_z^z = 
     &	\frac{1}{2x^2}\left[ P_1'^2 + \frac{P_2'^2}{g_2^2} \right] - \frac{\varepsilon_2}{g_2} 
\frac{P_1' P_2'}{x^2} \\
     & 	+\left[f_1' + \frac{P_1^2 f_1^2}{x^2}\right]
	+q_2^2\left[f_2' + \frac{P_2^2 f_2^2}{x^2}\right] 
     	- q_3^2\left[\bar{\chi}^2 f_3^2 -  f_3' \right] \\
     &	+\frac{\beta_1}{4}  \left(f_1^2-1\right)^2
	+q_2^4\frac{\beta_2}{4}  \left(f_2^2-1\right)^2
	+q_3^4\frac{\beta_3}{4}  \left(f_3^2 - 2\right)f_3^2 \\
     &	+q_2^2 q_3^2\beta_4  f_2^2 f_3^2  \ , \\
		\end{split}
\end{equation}
respectively, where $\bar{\chi}^2=\chi^2$ for $k^2 > \omega^2$ and
$\bar{\chi}^2=-\chi^2$ for $\omega^2 > k^2$. 
According to the formalism developed in \cite{carter2} the propagation velocities 
of tangential and longitudinal perturbations on superconducting strings are
\begin{equation}
\label{stabil}
c_T^2 = \frac{T}{E}   \ \ \ , \ \ \   c_L^2 = - \frac{dT}{dE}
\end{equation}
and stable string configurations require $c_T^2 > 0$ and $c_L^2 > 0$. 
In the following we will be interested in these
quantities and we will explore how $c_T^2$ and $c_L^2$ 
change when coupling the superconducting string to an Abelian-Higgs string
via a gauge field interaction. In \cite{PP} it was found that in the U(1)
$_{\rm local}$ $\times$U(1)$_{\rm global}$ model $c_T^2$ and $c_L^2$ are smaller than one
and that $c_T^2 > c_L^2$ such that longitudinal perturbations move slower than transverse ones.

Note that the energy-momentum tensor possesses an off-diagonal component 
$T_t^z$. However, since $T_t^z$ is proportional to the integral of $\omega k f_3^2$ we can always
choose a reference frame where $T_t^z$ vanishes. This can e.g. not be done when
two coupled currents are present on the string and the energy-momentum
tensor has to be diagonalized accordingly \cite{peter_martin}.


\subsection{Equations and boundary conditions}
The equations of motion result from the variation of the action (\ref{action}) with respect to the gauge and scalar fields. These read
\begin{align}
   &	2(x f_1')' = \beta_1 x (f_1^2 - 1) f_1 + 2\frac{P_1^2 f_1}{x} 
\label{EL1} \ , \\
   &	2(x f_2')' = \beta_2 q_2^2 x (f_2^2 - 1) f_2 + 2 \beta_4 q_3^2 x f_3^2 f_2 
+ 2\frac{P_2^2 f_2}{x}  \ , \label{EL2} \\
   &	2(x f_3')' = \beta_3 q_3^2 x (f_3^2 - 1) f_3 + 2 \beta_4 q_2^2 x f_3 f_2^2 + 2\chi^2 x f_3 \label{EL3} 
\end{align}
for the scalar field functions and 
\begin{align}
   &	\left(1 - \varepsilon_2^2 \right )x\left(\frac{P_1'}{x}\right)' =  
2 P_1 f_1^2 + 2\varepsilon_2 g_2 q_2^2 P_2 f_2^2  \ ,  \label{EL4} \\
   &	\left(1 - \varepsilon_2^2 \right) x \left(\frac{P_2'}{x}\right)' = 2 g_2^2 q_2^2 P_2 f_2^2 
+ 2 \varepsilon_2 g_2 P_1 f_1^2   \  \label{EL5} \\
 \end{align}
for the gauge field functions, respectively.
This system of coupled, ordinary differential equations needs to be solved numerically subject to appropriate boundary conditions.
The requirement of the regularity at $x=0$ leads to the following conditions
\begin{equation}
\label{BCS1}
   f_1(0) = 0 \ , \  f_2(0) = 0 \ , \ f_3'(0) = 0 \ , \ P_1(0) = n_1 \ , 
\  P_2(0) = n_2 \  \ . 
\end{equation}

Moreover, we want solutions that have finite energy per unit length. This leads to the following conditions
at $x=\infty$
\begin{equation}
 \label{BC2}
  f_1(\infty) = 1 \ , \  f_2(\infty) = 1 \ , \  f_3(\infty) = 0  \ , \  P_1(\infty) = 0 
\ , \  P_2(\infty) = 0 
 \ .
\end{equation}
Note that the trivial solution $f_3(x) \equiv 0$ is a solution
to the equations of motion fulfilling the above boundary conditions. In order to
force the $f_3(x)$ to be non-trivial, we impose a further boundary condition $f_3(0)=\gamma\neq 0$.
This in turn overdetermines the problem or to state it differently the choice
of $\gamma$ will fix the value of $\chi$. Practically, we will make $\chi$ a 
function $\chi(x)$ and implement 
a differential equation for $\chi$ which is of the form $\chi''=0$. The numerical
program will then determine $\chi$ for a given value of $\gamma$.


Finally, to ensure that the  U(1) symmetry associated to the fields $f_2$, $P_2$ is broken and 
the U(1) symmetry
associated to the field $f_3$ remains unbroken we have to fulfill a number of conditions for the coupling constants
in the model. To understand these conditions let us look at the potential associated to the scalar fields of the superconducting string which in rescaled
variables reads
\begin{equation}
\mathcal{V}_{23}(f_2^2,f_3^2)=
\label{pot}
q_2^4\frac{\beta_2}{4}  \left(f_2^2-1\right)^2
+q_3^4\frac{\beta_3}{4}  \left(f_3^2 - 2\right)f_3^2 \\
     +q_2^2 q_3^2\beta_4  f_2^2 f_3^2  \ .
\end{equation}
The conditions then are \cite{BHBC}
\begin{equation}
\label{IA}
\beta_3 q_3^4 < \beta_2 q_2^4  \ ,
\end{equation}

\begin{equation}
\label{IIA}
\beta_3 q_3^2  < 2 \beta_4 q_2^2  \ ,
\end{equation}

\begin{equation}
\label{IIIA}
\sqrt{\beta_2 \beta_3} < 2 \beta_4   \ ,
\end{equation}

\begin{equation}\label{IVA}	2 C \frac{\beta_2 \beta_4}{\beta_3^2} < \frac{q_3^4}{q_2^4}   \ .
\end{equation}
Condition \eqref{IA} is given by the inequality between the minima: 
$\mathcal{V}_{23}(1,0) < \mathcal{V}_{23}(0,1)$ when $x \rightarrow \infty$, 
such that the system chooses the former. Condition \eqref{IIA} is given by 
the requirement that the mass of $f_3(x)$ given in rescaled quantities by
$m_{\rm S}^2=2\beta_4 q_2^2 - \beta_3 q_3^2$ is positive. Condition \eqref{IIIA} is a 
restrictive lower bound which ensures that both  $\mathcal{V}_{23}(1,0)$ and $\mathcal{V}_{23}(0,1)$ are local minima.
Condition \eqref{IVA} was found numerically \cite{BHBC} with $C \approx 1.4$ and makes the condensate stable.
When $\chi \neq 0$ more general symmetry breaking conditions were found by replacing 
the potential in equation \eqref{pot} with an effective potential, as described in \cite{LS}. These conditions read:

\begin{equation}
\label{IB}
\beta_3\left(q_3^2-2\frac{\chi^2}{\beta_3}\right)^2 < \beta_2 q_2^4  \ ,
\end{equation}

\begin{equation}
\label{IIB}
\beta_3 q_3^2 - 2 \chi^2 < 2 \beta_4 q_2^2  \ ,
\end{equation}

\begin{equation}
\label{VB}
\chi^2 < \frac{\beta_3 q_3^2}{2} - \sqrt{\beta_2 \beta_4} q_2^2  \ .
\end{equation}
Thus, condition \eqref{IB} is the generalization of \eqref{IA}, while \eqref{IIB} is 
the generalization of \eqref{IIA} ensuring that the current
does not ``create'' $\phi_3$-particles. Also, \eqref{VB} is a condensate stability condition, 
found from a theoretical analysis of small perturbations in the field $|\phi_3|$ within 
the effective potential by considering the analogue of a quantum harmonic oscillator.
It is interesting to note here that, when going back to $\chi^2 \rightarrow 0$, \eqref{VB} 
reduces to \eqref{IVA} for $C=2$. This means that restriction 
\eqref{IVA} had been in the literature, in a slightly modified
and generalized version and was only rediscovered by \cite{BHBC} through 
a different approach. Note that condition \eqref{IB} together with condition \eqref{IIIA}
leads to the condition discussed in \cite{PP}, the so-called phase frequency threshold condition
beyond which no stationary solutions exist and energy and tension diverge.

\section{Numerical results}
\label{results}

This section presents the numerical results obtained using a collocation method \cite{colsys} to solve the system of coupled
ordinary differential equations subject to the appropriate boundary conditions. The parameters defining the potential are kept constant for the entire study:
\begin{equation}
	\label{PotParams}
   	\beta_1 = 2.0, ~~ \beta_2 = 2.0, ~~ \beta_3 = 80.0, ~~ \beta_4 = 10.0,~~ q_2 = 1.0, ~~ q_3 = 0.37
\end{equation}
such that they simultaneously satisfy, with reasonable margins, \eqref{IA} -- \eqref{VB} 
in the $\chi^2 = 0$ limit. Furthermore, the current was also restricted by the validity 
of \eqref{IB}, \eqref{IIB} for all $\chi^2$.
 
In order to investigate the stability of our solutions we will show tension-energy characteristics $T(E)$. 
The current is controlled by the value $f_3(0)\equiv\gamma$.
In Fig.\ref{Generic:1} we show how the quadratic norm $\chi^2$ principally depends on the value of the condensate $\gamma$ for such strings. 
In Fig.\ref{Generic:2} we show how the tension $T$ depends in principle on the energy $E$. 
In a generic way, we shall refer to these diagrams as current nature diagram and tension-energy diagram, respectively. 
Note that all tension-energy diagrams include the contribution of the Abelian-Higgs string, 
even when this is completely decoupled from the superconducting string, i.e. for 
$\varepsilon_2=0$. 
In this case, this contribution only shifts the values of $E$ and $T$ by a constant value.

\begin{figure}[!h]
\centering
	\subfloat[]{\includegraphics[width=7.8cm]{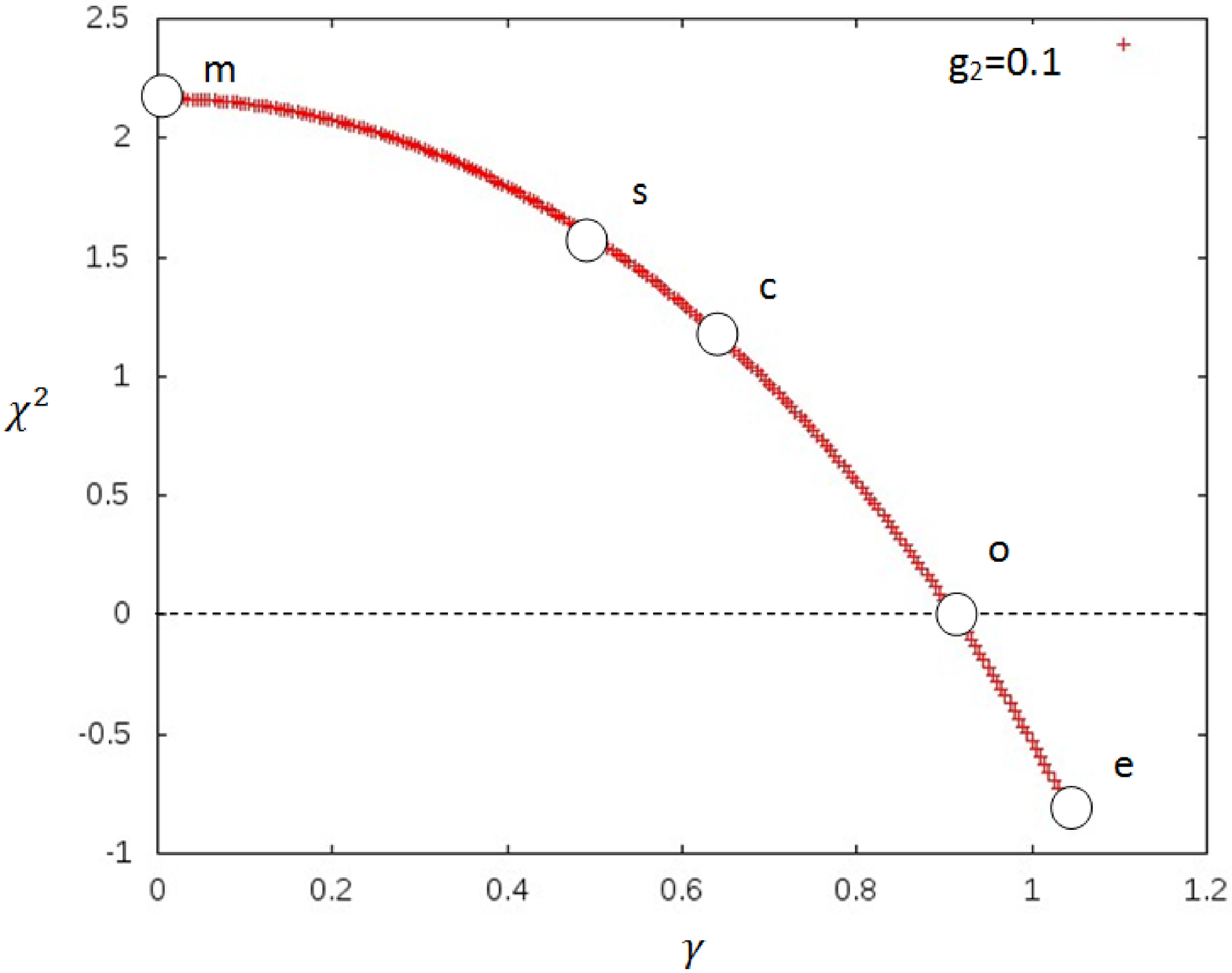}\label{Generic:1}} 
	\subfloat[]{\includegraphics[width=8.2cm]{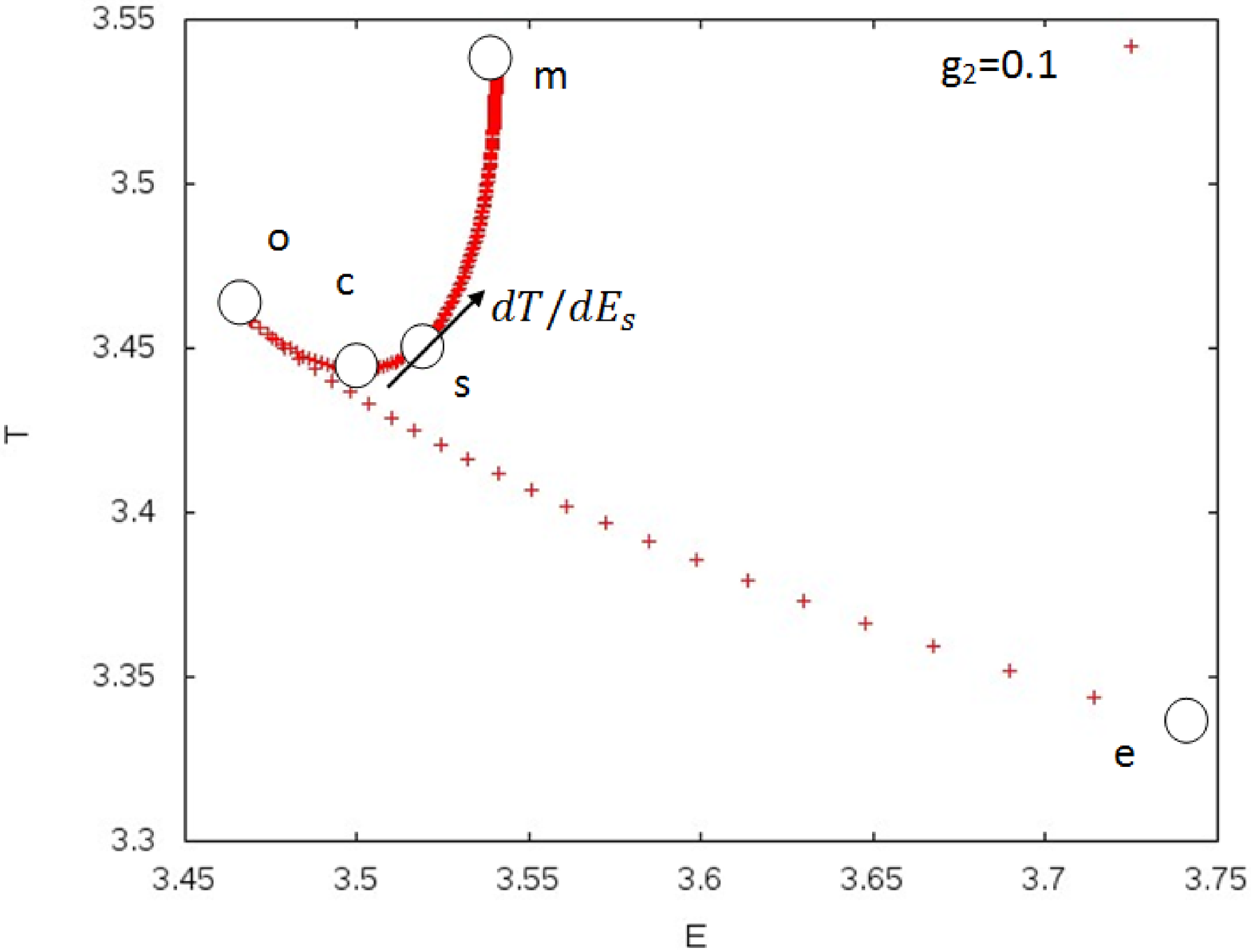}\label{Generic:2}}
	\caption{We show the principal dependence of $\chi^2$ on the value of the condensate $f_3(0)=\gamma$ \subref{Generic:1} 
and the principal dependence of the tension $T$ on the energy $E$ \subref{Generic:2}.}
\label{Generic}
\end{figure}

For a convenient future use, we define five important points, or ``current markers'', 
in these diagrams: point ``e'' denotes the lower limit of the electric 
region, 
where $\chi^2$ attains its lowest, negative value, 
just before falsifying condition \eqref{IB}. This corresponds to the phase frequency
threshold discussed in \cite{PP} and marks the value of $\chi^2$ beyond which no stationary
solutions exist with the energy $E$ and tension $T$ diverging there. Note that the solutions
are stable in the full electric regime with $T$ and $E$ positive and $\frac{dT}{dE} < 0$
between points ``o'' and ``e''. 
Point ``o'' marks the transition between the electric and magnetic regions, 
where $\chi^2=0$. ``c'' is the critical point in the magnetic region where $T(E)$ 
attains its minimum, with $\frac{dT}{dE}=0$, at the transition 
between the stable (o-c, $\frac{dT}{dE} < 0$) and unstable (c-m, $\frac{dT}{dE} > 0$) magnetic regions.
``m'' gives the upper limit of the magnetic region. For $\chi^2$ larger than the value
at ``m'' the current carrier field decouples from the Higgs field of the cosmic
string and we are left with a simple Abelian-Higgs string coupled to another Abelian-Higgs string.
Finally, ``s'' is the point where condition \eqref{VB} is falsified 
-- region s-m is thus also unstable with respect to \eqref{VB}. 
These points divide the diagrams into multiple ``current regions''. 
In this line of thought, we define the following 
energy differences: $E_{e-o}:=E_e - E_o$ -- energy width of the electric region; $E_{m-o}:=E_m-E_o$ -- 
energy width of the magnetic region; $E_{c-o}=E_c-E_o$ -- energy width of the stable, 
magnetic region (contained within $E_{m-o}$); $E_{s-c} = E_s - E_c$ -- energy width between ``s'' and ``c'' defined above. 
The definitions of these four current regions are essential for evaluating 
how the main features of the current evolve when changing various parameters, 
and will be extensively used below. We also use the derivative $\left(\frac{dT}{dE}\right)_{s}$ as a measure of the instability of point ``s'' with respect to \eqref{stabil}.                                       

We start by mentioning some limiting cases of our model, which are equivalent 
to those used in former studies -- see 
Section \ref{LimCases}.  
We then proceed by 
investigating the effects of the gauge interactions on the current. 
and present a detailed analysis of the $\varepsilon_2$-mediated interaction 
between the broken symmetry of the Abelian-Higgs string and the broken symmetry of the 
superconducting string -- see Section \ref{I2}.

For convenient future use, we choose to denote the three sectors of our field theoretical model in the following 
way: ``sector 1'' -- the broken symmetry of the Abelian-Higgs (dark) string; 
``sector 2'' -- the broken symmetry of the superconducting string; ``sector 3'' 
-- the unbroken symmetry of the superconducting string. 
The labeling of our functions: $f_1$, $f_2$, $f_3$, $P_1$, $P_2$ is consistent with this notation. Furthermore, note that the values of $E$, $T$, $E_{e-o}$, $E_{m-o}$, $E_{c-o}$ and $E_{s-c}$ are given in units of $2\pi$ in all plots. We use $E_{A-B}$ as a generic notation for the above mentioned energy widths of the four current regions.

\subsection{Limiting cases}
\label{LimCases}
We have checked the consistency of our numerical techniques by going to certain limiting cases, 
which have been studied before. We have checked the known results and found that they agree (for more details see 
\cite{alex_thesis}).
In \cite{BHFA} the model introduced above was studied for $\beta_4=0$ and for vanishing
current $f_3(x)\equiv 0$ describing the (attractive) interaction between a dark string and a cosmic
string. In this limiting case, this model can also be seen as an alternative field theoretical
model to describe (p,q)-strings. In the original proposal \cite{saffin} the two Abelian-Higgs strings
had been coupled via a potential interaction term, while here the interaction is
mediated via the gauge fields. 
The $\varepsilon_2=0$ limit corresponds to an Abelian-Higgs string and a superconducting
string that do not interact. Superconducting strings in a 
U(1)$_{\rm local}$ $\times$U(1)$_{\rm global}$ model have been studied 
e.g. in \cite{Witten,PP,BHBC}.

\subsection{Superconducting strings interacting with cosmic strings $\varepsilon_2\neq 0$}
\label{I2}

This section deals with the gauge interaction between sector 1 and 2, 
which is mediated by the parameter $\varepsilon_2$ coupling the two broken symmetries of our model. 
In order to understand the influence of $\varepsilon_2$ on the current the variation of the tension-energy diagram 
is studied using the current markers and regions as defined above. 
The parameters of the potentials are kept constant to \eqref{PotParams}. In Fig.\ref{AllProfiles2} we show the profiles of a 
typical solution in this case.

\begin{figure}[!h]
\centering
\includegraphics[width=10cm]{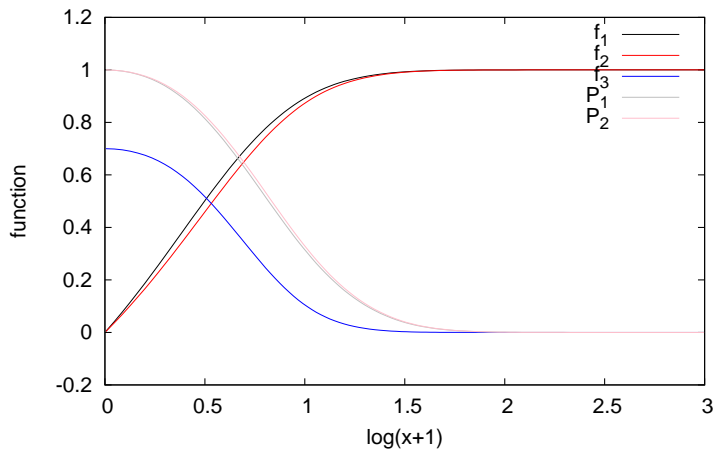}
\caption{We show the profiles of the functions describing a superconducting 
string interacting with a cosmic string 
with $g_2=1.0$, $\varepsilon_2=0.01$, $n_1 = 1$, $n_2 = 1$, $\gamma = 0.7 ~ 
(\chi^2 \approx - 0.50)$.}
\label{AllProfiles2}
\end{figure}  

In Section \ref{I2-g2} we will give our results for the behaviour of the current when changing 
the gauge coupling constant $g_2$, while keeping $\varepsilon_2 = 0$. Then, for several
 values of this parameter around $g_2=1$ we will demonstrate in Section \ref{I2-eps2} 
how the current changes when changing the gauge interaction parameter $\varepsilon_2$. 
For both cases, the windings are kept to their lowest possible values 
$n_1=1$, $n_2 =1$.
Finally, the gauge constant is set to $g_2 = 1$ in Section \ref{I2-wind}, 
and the influence of the windings $n_1$ and $n_2$ is studied.

\subsubsection{Effect of gauge coupling constant $g_2$}
\label{I2-g2}

\begin{figure}[!h]
\centering
	\subfloat[]{\includegraphics[width=8cm]{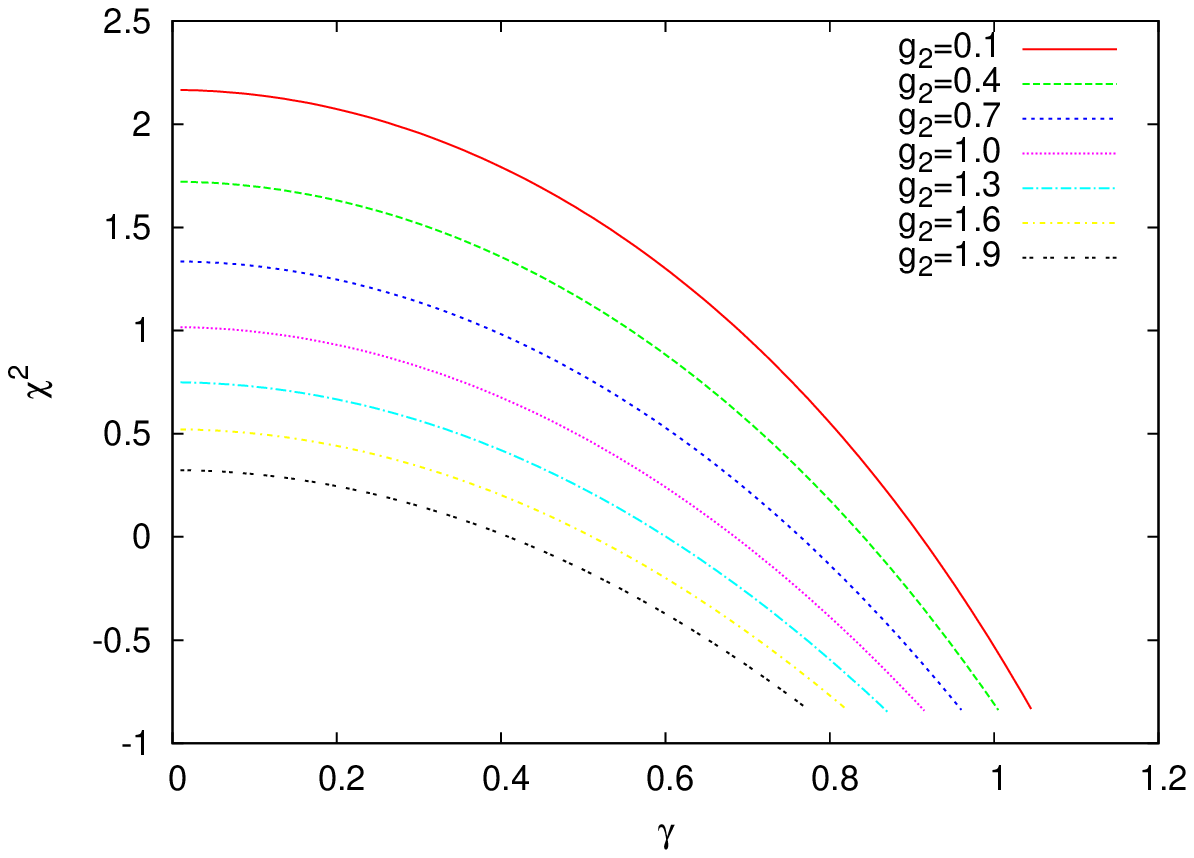}\label{gbA:1}} 
	\subfloat[]{\includegraphics[width=8cm]{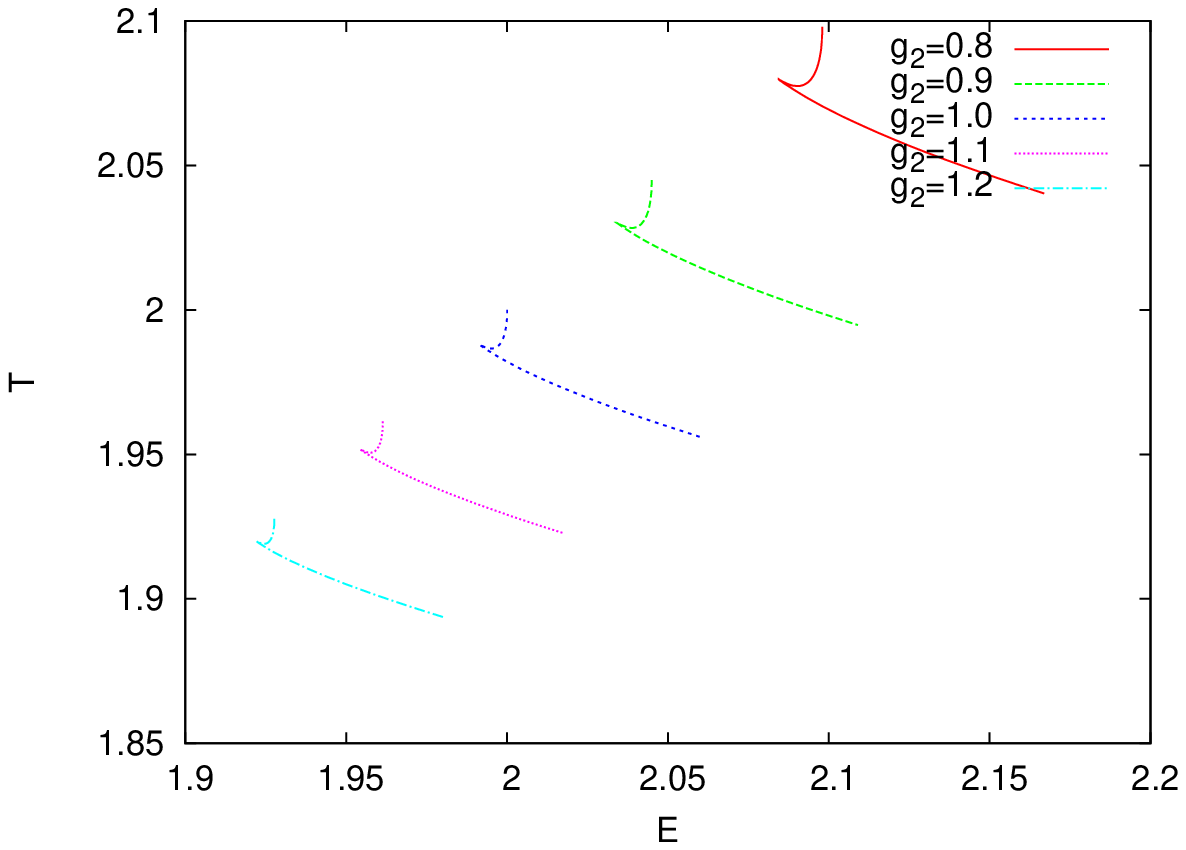}\label{gbA:2}}
	\caption{We show the influence of $g_2$ 
on $\chi^2$ \subref{gbA:1} and on $T(E)$ \subref{gbA:2} for $\varepsilon_2=0$ and $n_1=n_2=1$.}
\label{gbA}
\end{figure}

First of all, we illustrate the influence of $g_2$ on the value of 
$\chi^2$ and on the tension-energy diagram $T(E)$ in Fig.\ref{gbA}. Note that
with our choice $q_2=1$ the value of $g_2$ is equal to the ratio between
the two gauge boson masses $m_{\rm W,1}$ and $m_{\rm W,2}$, $g_2=m_{\rm W,2}/m_{\rm W,1}$ 
and as such also equal to the ratio between the flux tube core widths: $g_2=\rho_{\rm W,1}/\rho_{\rm W,2}$.
From Fig.\ref{gbA:1} it is obvious that an increase in $g_2$ leads to the decrease of
the maximal possible current on the string (at $\gamma=0$) before it decouples and leaves
behind a standard cosmic string without current. Moreover, the value 
of $\gamma$ beyond which no stationary solutions exist -- the so-called phase frequency
threshold which is at $\chi_e^2 \approx - 0.848$ for our choice of parameters (\ref{PotParams}) 
-- also decreases. 
The energy and tension decrease when $g_2$ increases, i.e. 
when decreasing the radius of the flux tube in sector 2  with respect to the radius 
of the flux tube in sector 1.
One can also see that the larger the coupling $g_2$ the more the electric region $\chi^2 \le 0$ 
is favoured over the magnetic one, as $\chi^2(\gamma)$ is 
shifted downwards which can also be seen in Fig.\ref{gbA:2}, 
where the size of the magnetic branch seems to decrease 
faster than the size of the electric branch. 
This non-trivial effect indicates that the width of the flux tube in 
sector 2 has a significant influence on the 
nature of the current in sector 3.

\begin{figure}[!h]
\centering
	\subfloat[$c_T^2(\chi^2)$ for several $g_2$]{\includegraphics[width=8cm]{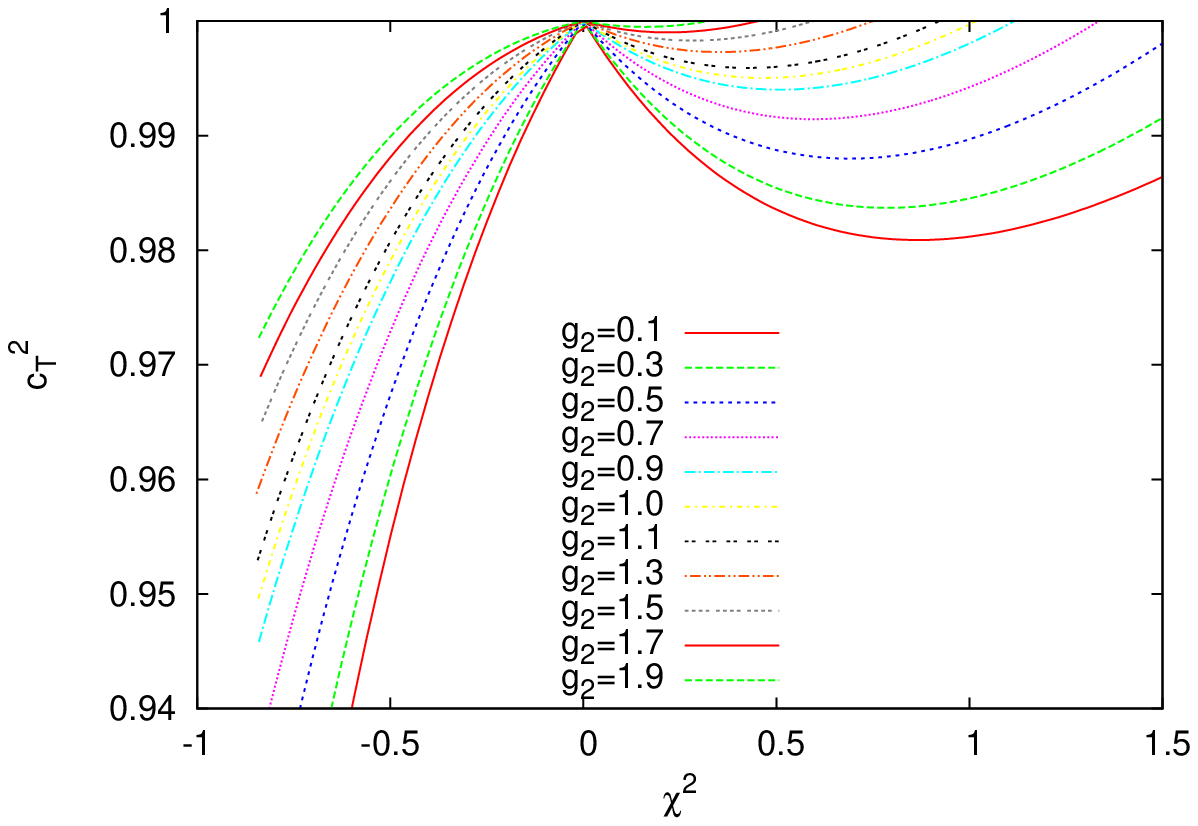}\label{Vels1:T}} 
	\subfloat[$c_L^2(\chi^2)$ for several $g_2$]{\includegraphics[width=8cm]{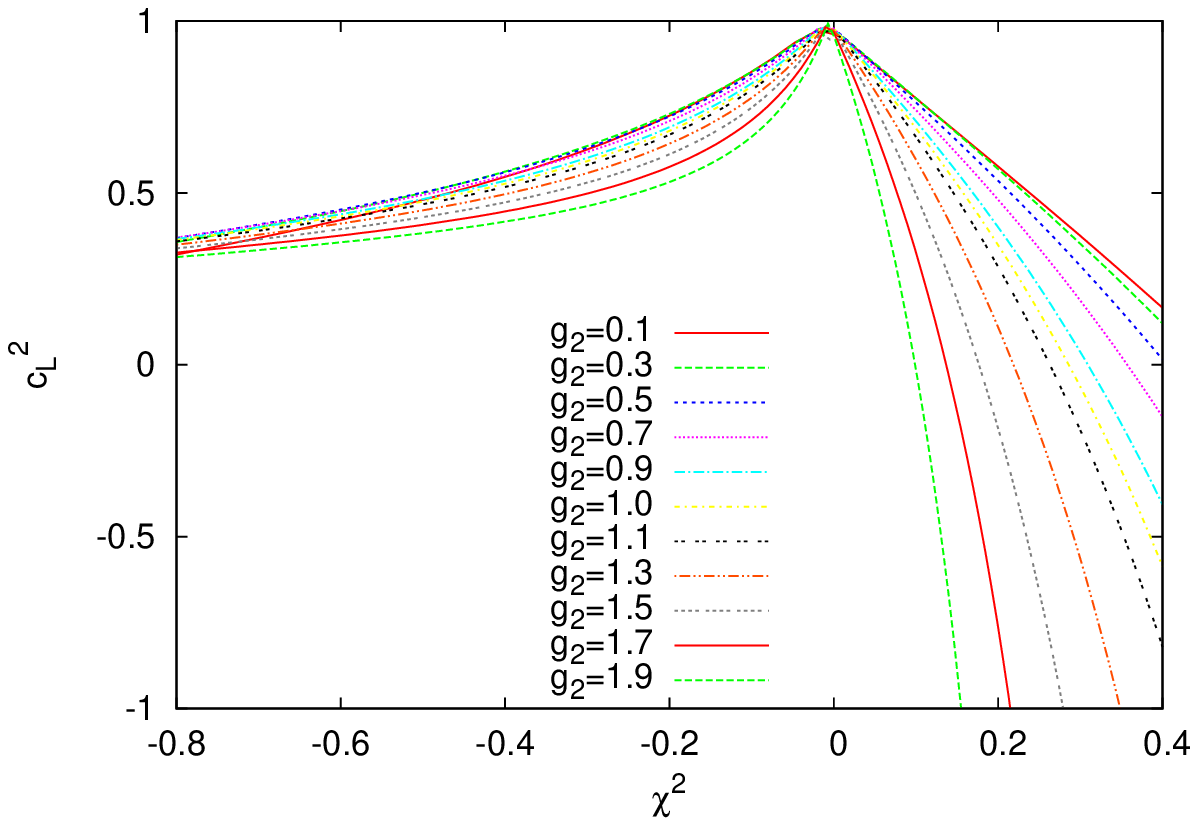}\label{Vels1:L}}
	\caption{We show the influence 
of $g_2$ on the tangential \subref{Vels1:T} and 
longitudinal \subref{Vels1:L} propagation velocities as functions of $\chi^2$ for 
$\varepsilon_2=0$ for $n_1=n_2=1$.}
\label{Vels1}
\end{figure}

We also illustrate the influence of $g_2$ on the tangential $c_T^2$ 
and longitudinal $c_L^2$ propagation velocities in Figs.
\ref{Vels1:T} and \ref{Vels1:L}, respectively. 
It is clear that $c_T^2$ increases with $g_2$ within both the 
electric and magnetic regions, except when $\chi^2 = 0$ (point ``o''),
 where this influence vanishes -- a simple consequence of the fact that $E = T$. 
This means that the decrease of the width of the flux tube of the superconducting
string with respect to that of the standard cosmic string leads to an increase
of the propagation velocity of the transverse perturbation. 
On the other hand, $c_L^2$ decreases with $g_2$ in 
the magnetic region and partly in the electric region 
-- a close look at Fig.\ref{Vels1:L} shows that, as $\chi^2<0$ decreases and 
the curves approach each other, some of them intersect. 
This means that in the magnetic region and partly in the electric region 
the difference between the transverse and longitudinal propagation speeds
becomes larger when decreasing the width of the flux
tube of the superconducting string.

Fig.\ref{Vels1:L}
also shows that the transition to the unstable, magnetic region (point ``c'') is attained for lower $\chi^2$ as $g_2$ increases. 

\begin{figure}[!h]
\centering
	\subfloat[$\chi^2(g_2)$ for all current markers]{\includegraphics[width=8cm]{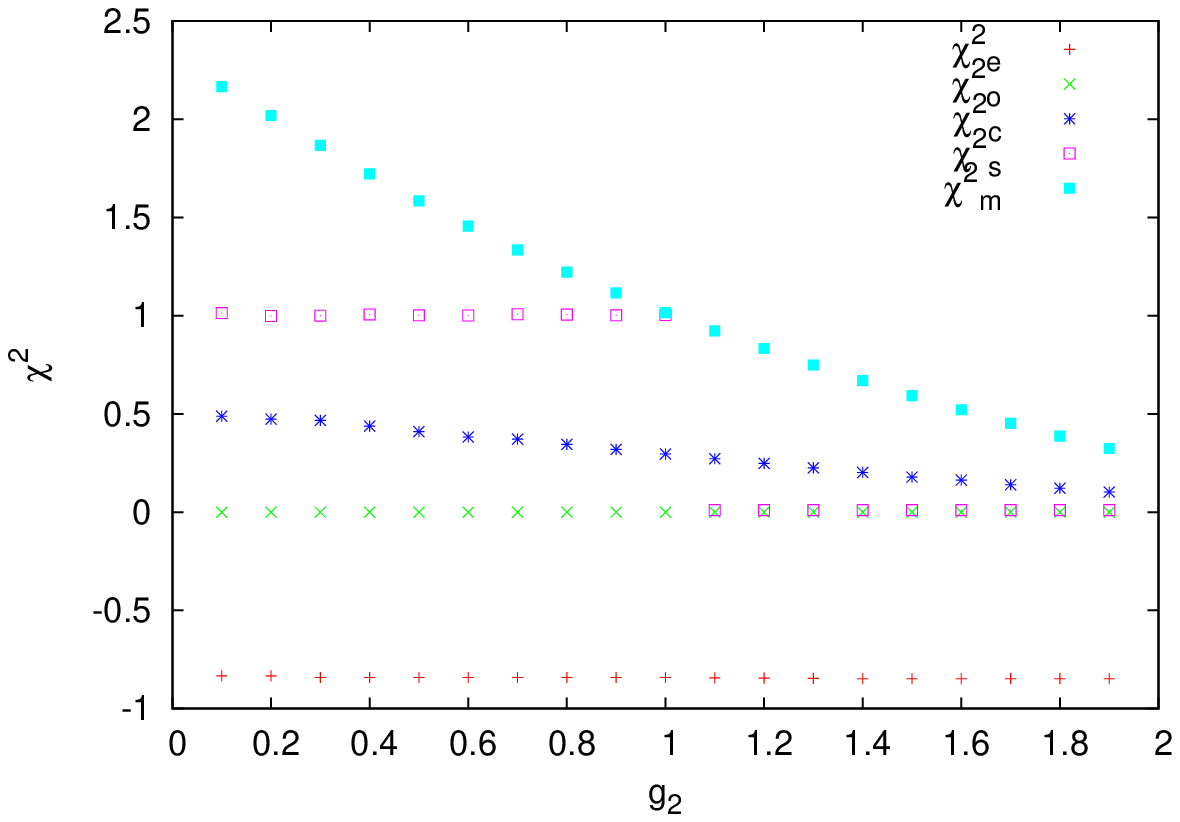}\label{gbB:1}} 
	\subfloat[$E_{A-B}(g_2)$ for all current regions]{\includegraphics[width=8cm]{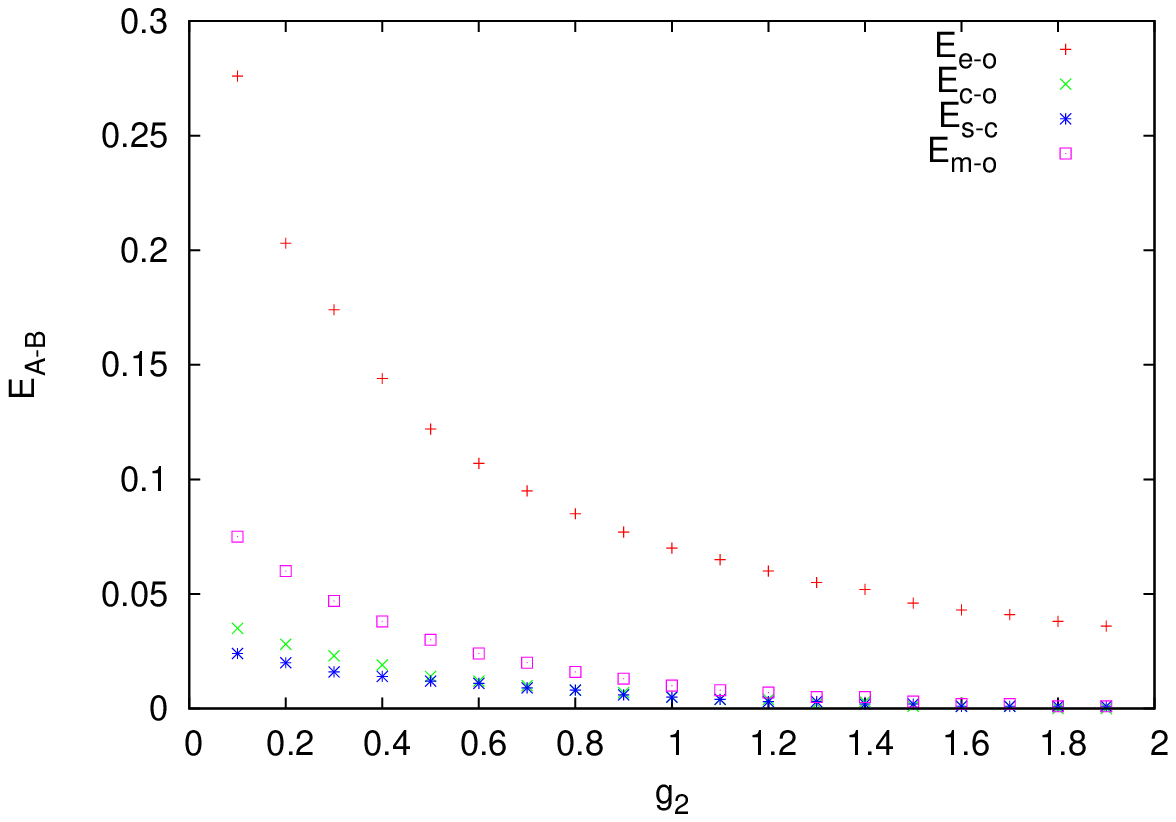}\label{gbB:2}}
	\caption{We show the current markers as functions 
of $g_2$ in $\chi^2$-space \subref{gbB:1} and in $E$-space \subref{gbB:2}
for $\varepsilon_2=0$ and $n_1=n_2=1$.}
\label{gbB}
\end{figure}

In order to better understand these effects, the value at all current markers and the 
energy intervals of all current regions are plotted as functions of $g_2$ in Fig.\ref{gbB:1} and Fig.\ref{gbB:2}, respectively.
Fig.\ref{gbB:1} shows that the complete magnetic and stable magnetic regions both 
decrease with $g_2$ in $\chi^2$ space, while the electric region 
remains the same (because both $\chi^2_e$ and $\chi^2_o$ are 
constant with $\chi^2_e\approx -0.848$ for our choice of parameters (\ref{PotParams}) and
$\chi^2_o=0$ by definition). 
The value of $\chi_s^2$ is constant (see \eqref{VB}), but joins the curve for $\chi_m^2$ at $g_2 = 1.0$. 
Fig.\ref{gbB:2} shows that the energy widths of all current regions decrease with $g_2$. It can be clearly seen that the magnetic region would vanish long before the electric one. 

\subsubsection{Effect of $\varepsilon_2$}
\label{I2-eps2}

\begin{figure}[!h]
\centering
	\subfloat[$g_2=1.0$.]{\includegraphics[width=8cm]{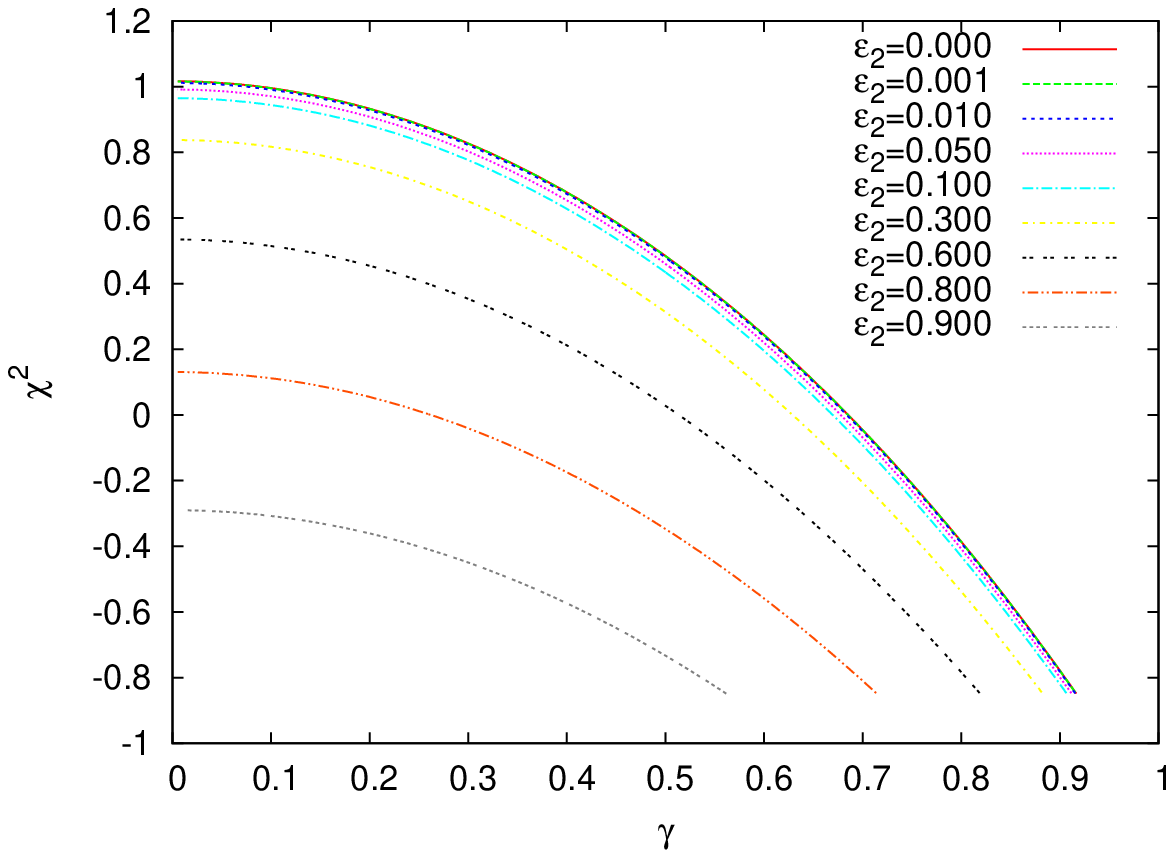}\label{ge:1}} 
	\subfloat[$g_2=1.0$]{\includegraphics[width=8cm]{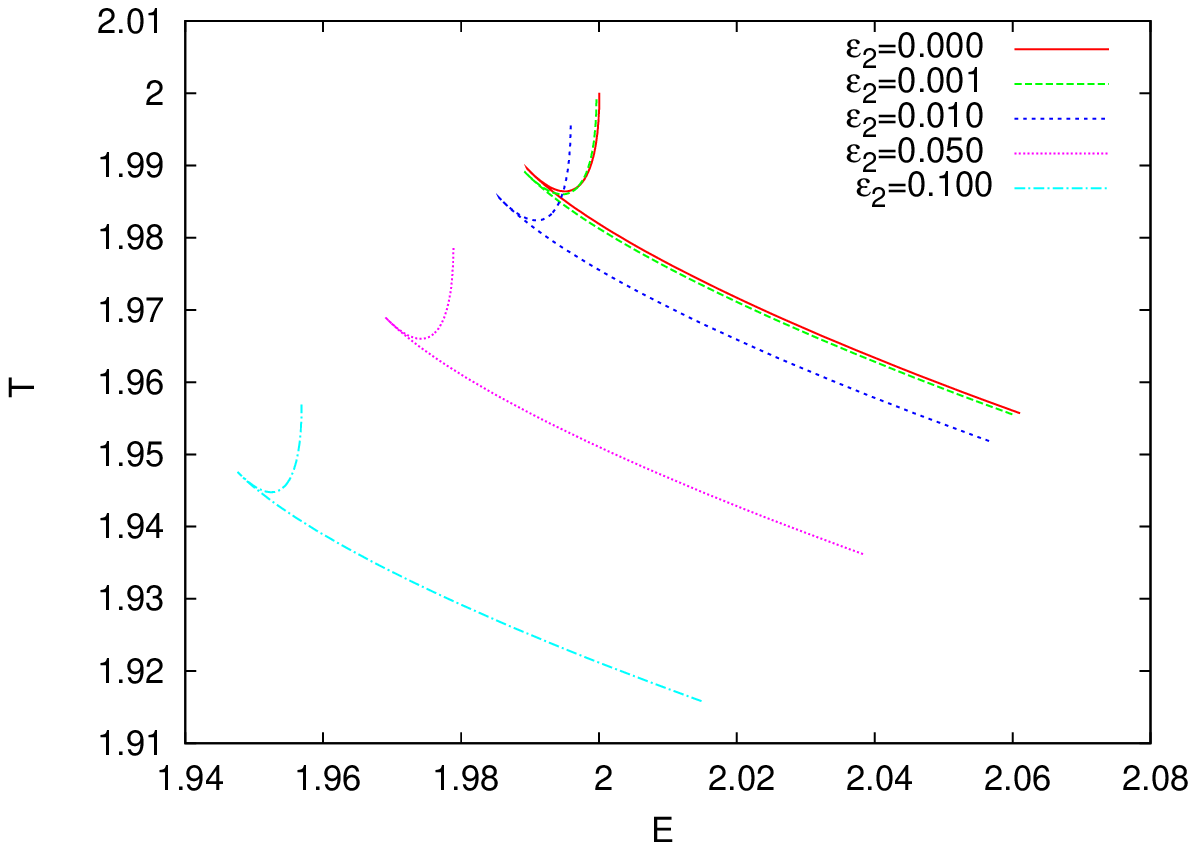}\label{ge:2}}
	\caption{We show the influence of $\varepsilon_2$ 
on $\chi^2$ \subref{ge:1} and on $T(E)$ \subref{ge:2} for $g_2=1$ and $n_1=n_2=1$.}
\label{ge}
\end{figure}

We illustrate the influence of $\varepsilon_2$ on the current nature $\chi^2(\gamma)$ 
and on the tension-energy diagram $T(E)$ in Fig.\ref{ge} for $g_2 = 1.0$. 
One can see that the influence is qualitatively similar to the influence of $g_2$.
Fig.\ref{ge:1} shows that the increase of $\varepsilon_2$ leads to the decrease
of the maximal possible current on the string before it decouples ($\gamma=0$ and hence $f_3(x)\equiv 0$).
The critical value of $\gamma$ beyond which no stationary solutions exist (the so-called
phase frequency threshold at $\chi^2_e\approx -0.848$) also decreases. For large enough $\varepsilon_2$ we find that
$\chi^2 < 0$ always and the magnetic regime completely disappears. We find that
this happens for $\varepsilon_2\approx 0.84$
if $g_2=1$. 

In Fig.\ref{ge:2} we show the dependence of $T$ on $E$ for reasonably small values of
$\varepsilon_2$. Increasing $\varepsilon_2$ decreases the tension $T$ and energy $E$ such
that the curves get shifted downwards along a line $T=E$ for increasing $\varepsilon_2$. 

\begin{figure}[!h]
\centering
	\subfloat[$c_T^2(\chi^2)$ for several $\varepsilon_2$]{\includegraphics[width=8cm]{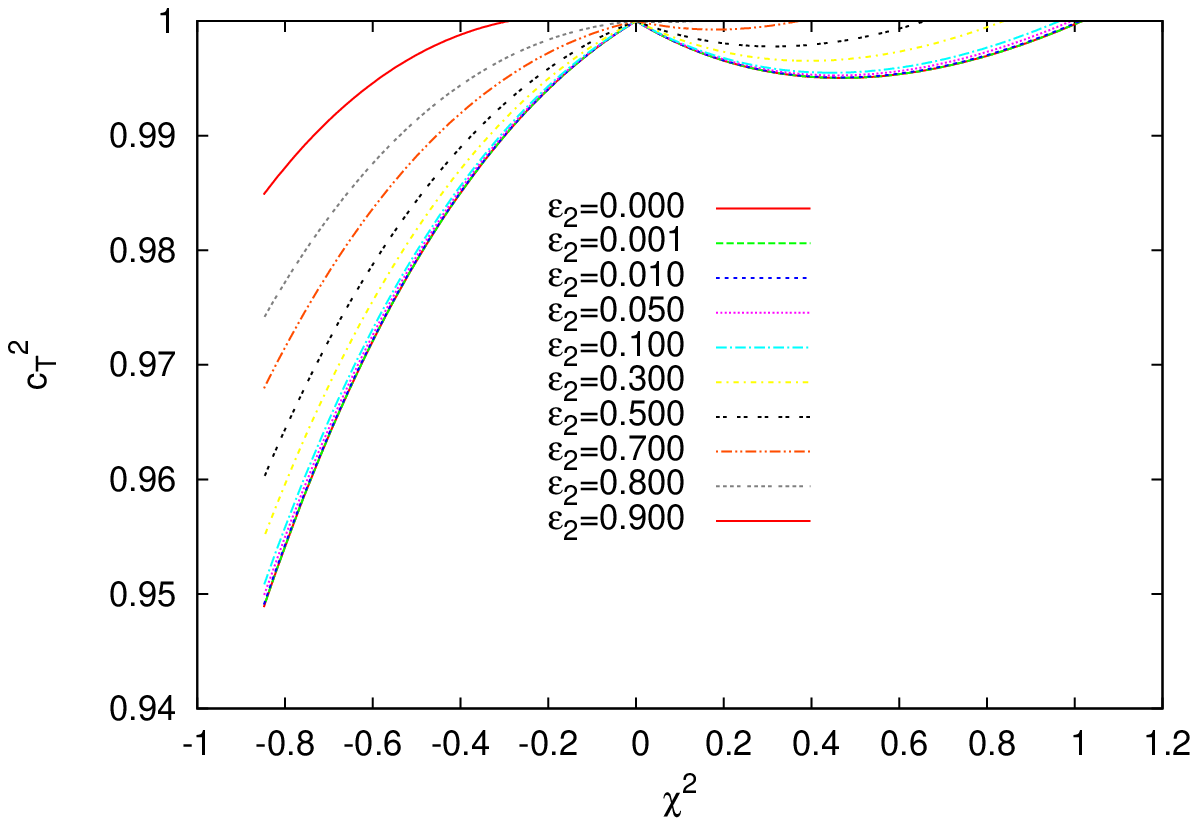}\label{Vels2:T}} 
	\subfloat[$c_L^2(\chi^2)$ for several $\varepsilon_2$]{\includegraphics[width=8cm]{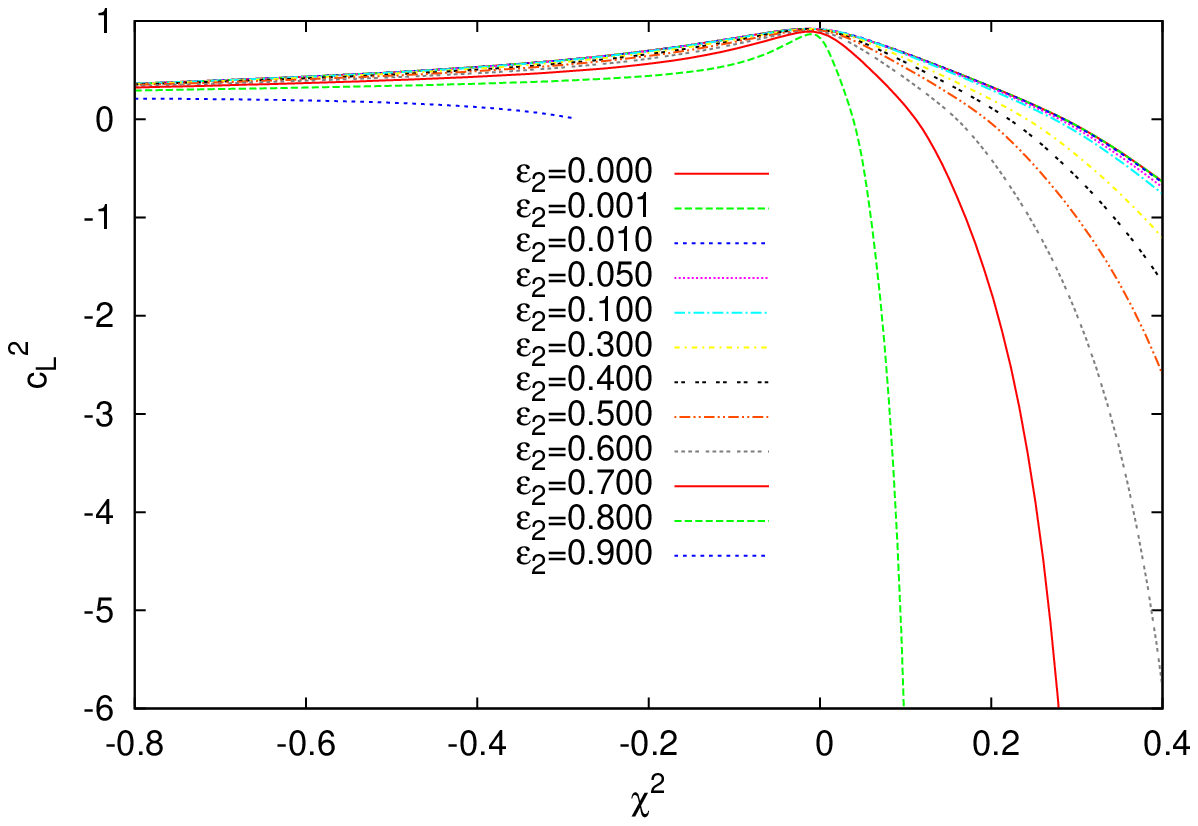}\label{Vels2:L}}
	\caption{We show the influence of $\varepsilon_2$ 
on the tangential \subref{Vels2:T} and longitudinal \subref{Vels2:L} 
propagation velocities as functions of $\chi^2$ 
for $g_2=0$ and $n_1=n_2=1$.}
\label{Vels2}
\end{figure}

We also illustrate the influence of $\varepsilon_2$ on the tangential $c_T^2$ 
and longitudinal $c_L^2$ propagation velocities in Fig.\ref{Vels2:T} and \ref{Vels2:L}, 
respectively. $c_T^2$ increases with $\varepsilon_2$ while 
$c_L^2$ decreases, within both the electric and magnetic regions. 
It is remarkable that, when $\varepsilon_2 = 0.9$ and the magnetic region 
is no longer present, the velocities $c_T^2 \rightarrow 1$ and $c_L^2 \rightarrow 0$ 
as the condensate vanishes $\gamma \rightarrow 0$ at $\chi^2 \approx -0.29$. In this case, the $T(E)$ curve only keeps part of the electric branch, as can be seen in Fig.\ref{ElBr}.

\begin{figure}[!h]
\centering
\includegraphics[width=8cm]{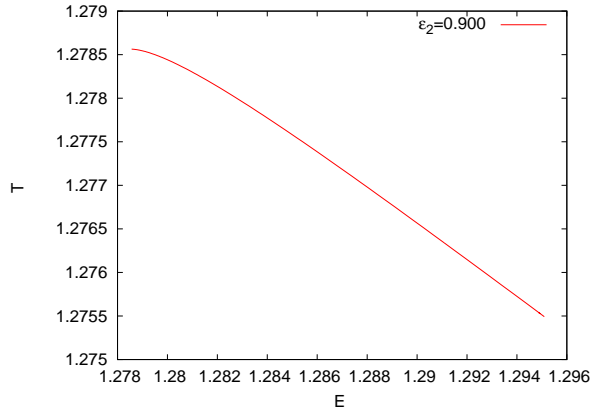}
\caption{We show the $T(E)$ diagram when the magnetic region is no longer present, for $g_2=1.0$, $\varepsilon_2=0.9$, $n_1 = 1$, $n_2 = 1$.}
\label{ElBr}
\end{figure}  

\begin{figure}[!h]
\centering
	\subfloat[]{\includegraphics[width=8cm]{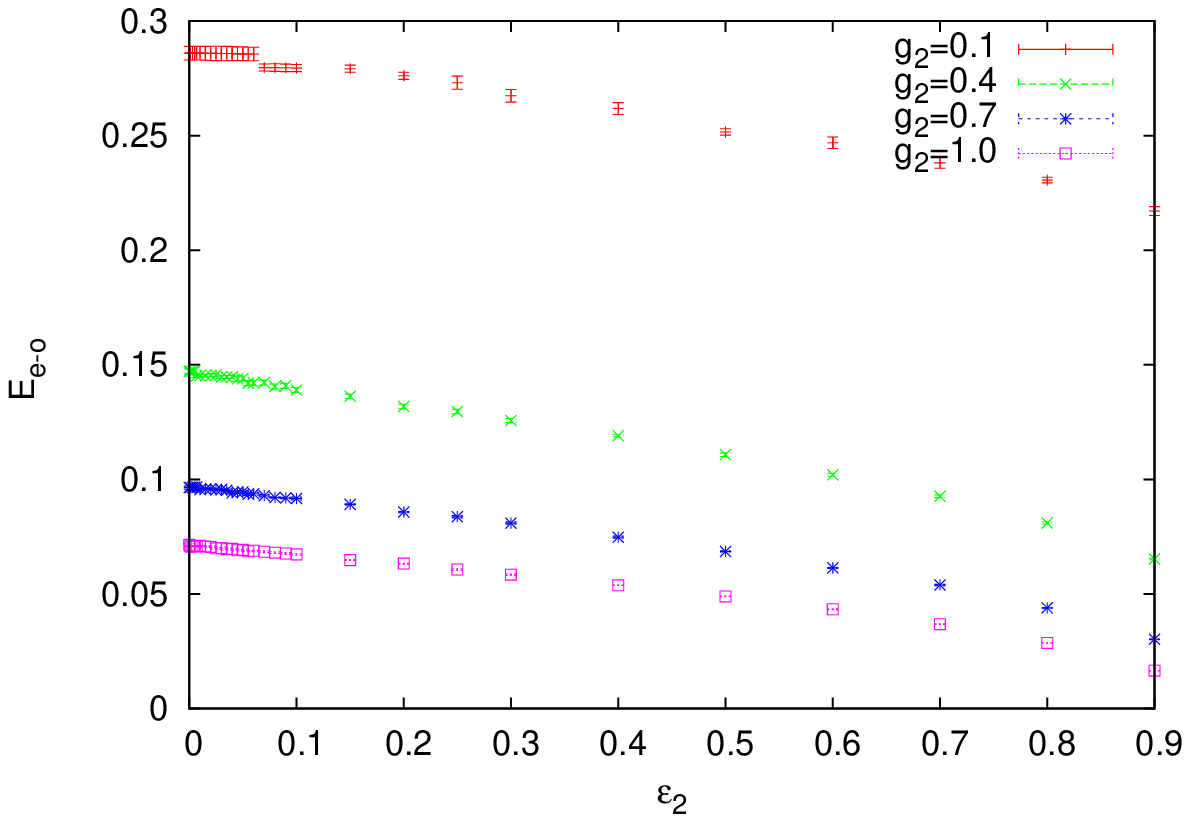}\label{eog:1}} 
	\subfloat[]{\includegraphics[width=8cm]{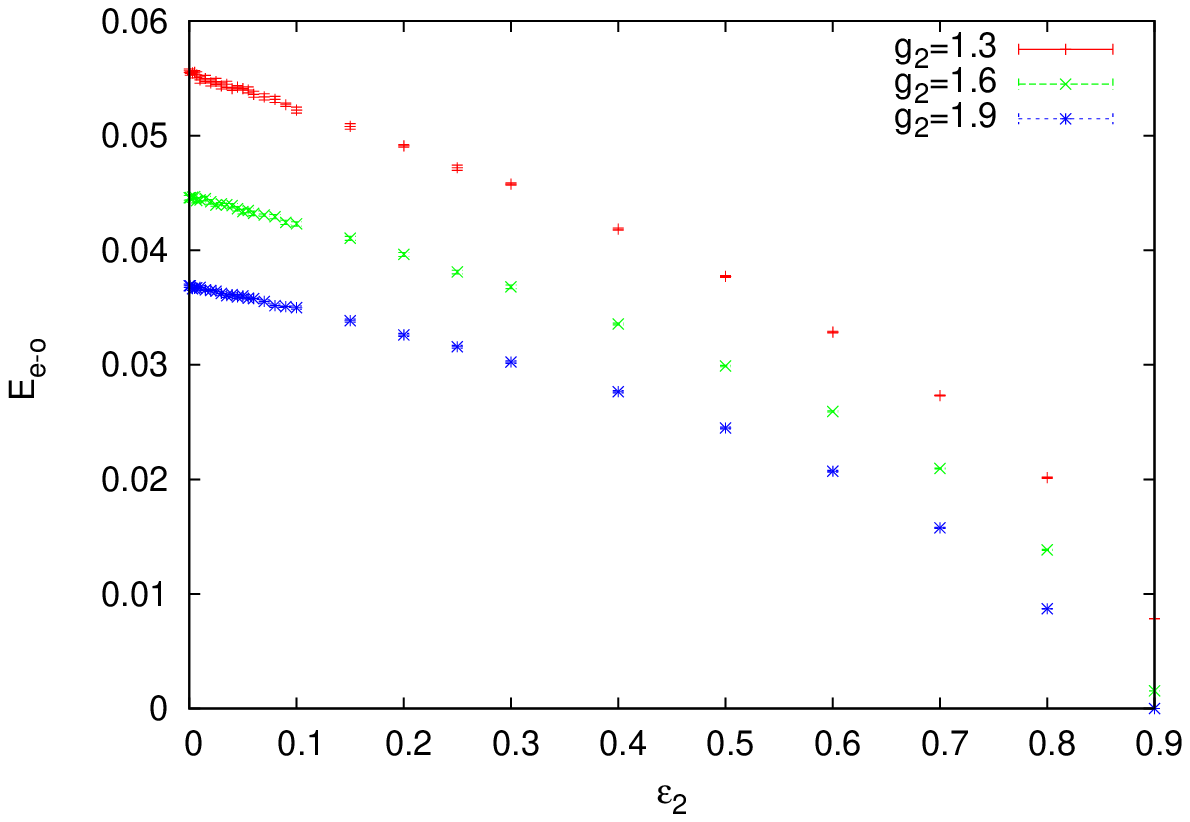}\label{eog:2}}
	\caption{We show the energy range of the electric region 
$E_{e-o}$ in dependence on $\varepsilon_2$ for several values of $g_2$}
\label{eog}
\end{figure}

We are mainly interested in how the energy widths 
of the main current regions change with $\varepsilon_2$. In Fig.\ref{eog}, Fig.\ref{eom}, Fig.\ref{eoc} and Fig.\ref{esc}
we show the values of $E_{e-o}$, $E_{m-o}$, $E_{c-o}$ and $E_{s-c}$, respectively
in dependence on $\varepsilon_2$ for several values of $g_2$. 
We observe that all values decrease with increasing $\varepsilon_2$. The quantitative form of these curves 
depends on  
$g_2$. A larger $g_2$ shifts the curves downwards 
correlating with a decrease of the width of the flux tube in sector 2.

\begin{figure}[!h]
\centering
	\subfloat[]{\includegraphics[width=8cm]{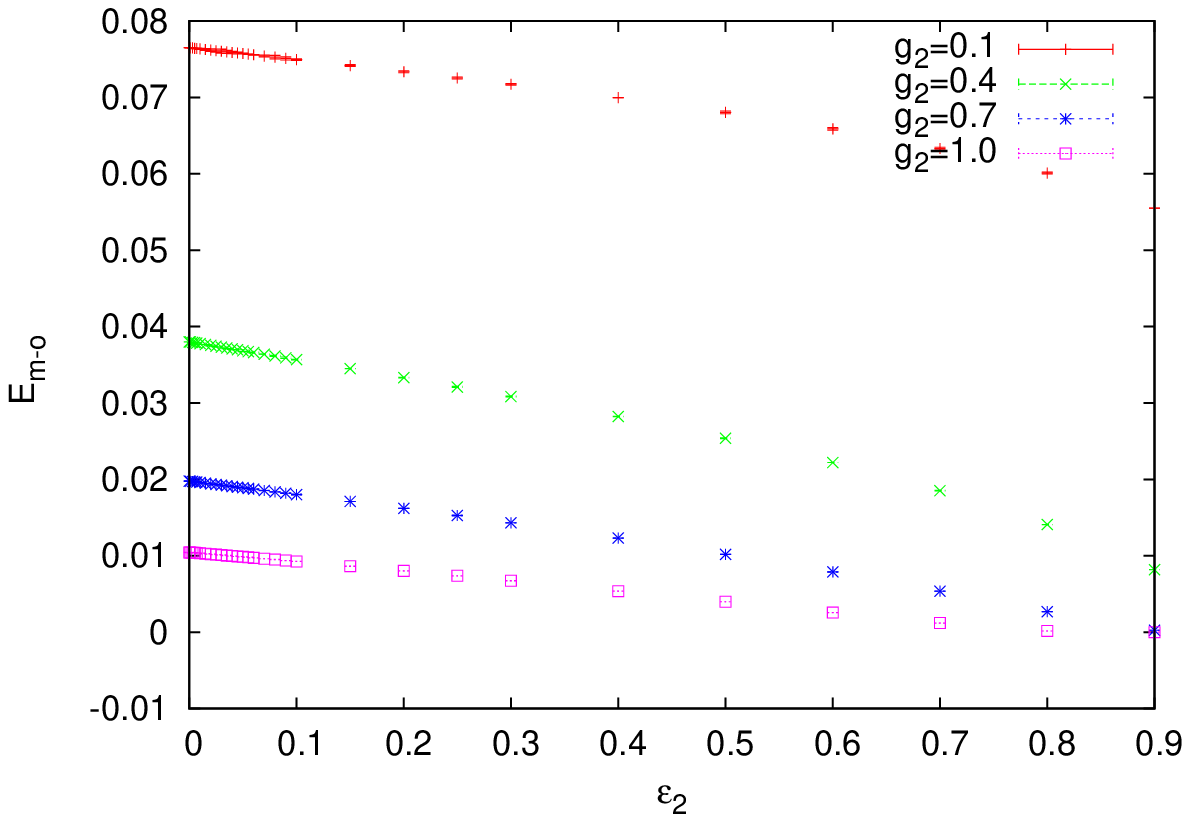}\label{eom:1}} 
	\subfloat[]{\includegraphics[width=8cm]{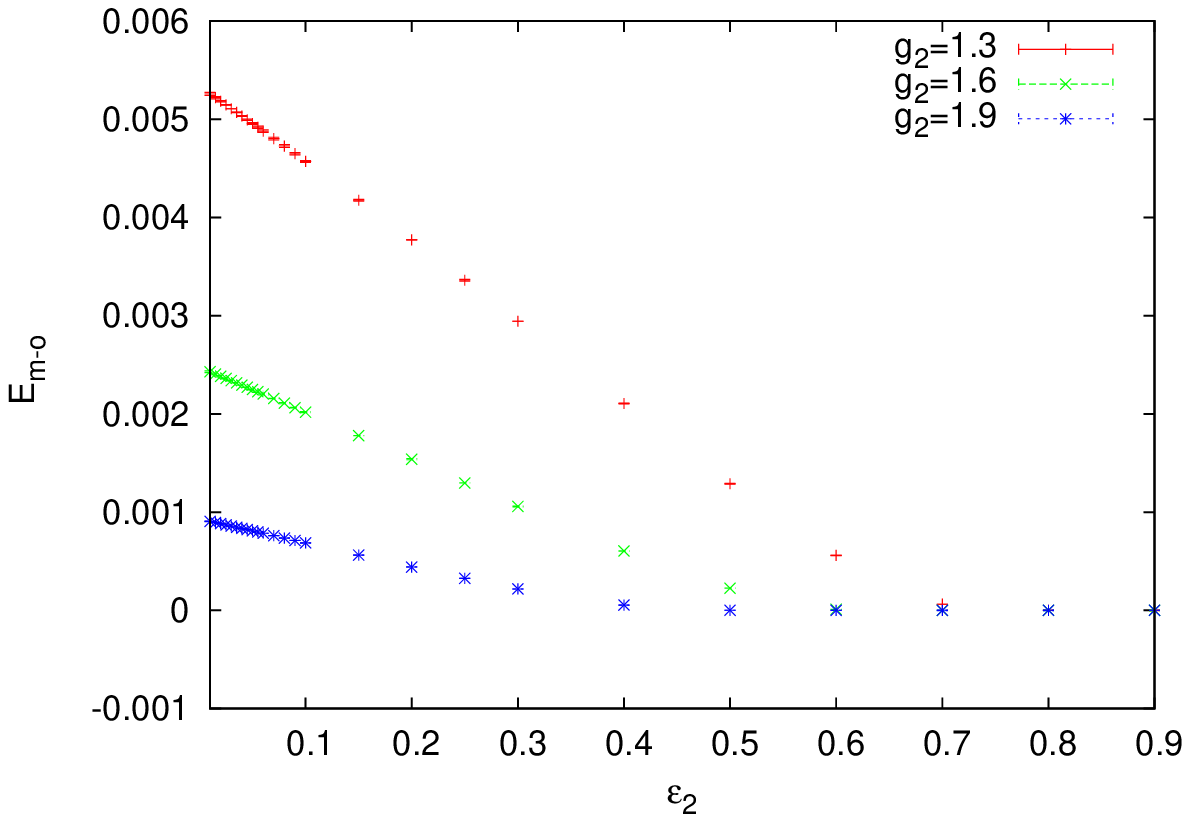}\label{eom:2}}
	\caption{We show the energy range of the magnetic region $E_{m-o}$ in dependence on 
	$\varepsilon_2$ for several values of $g_2$ and $n_1=n_2=1$.}
\label{eom}
\end{figure}

\begin{figure}[!h]
\centering
	\subfloat[]{\includegraphics[width=8cm]{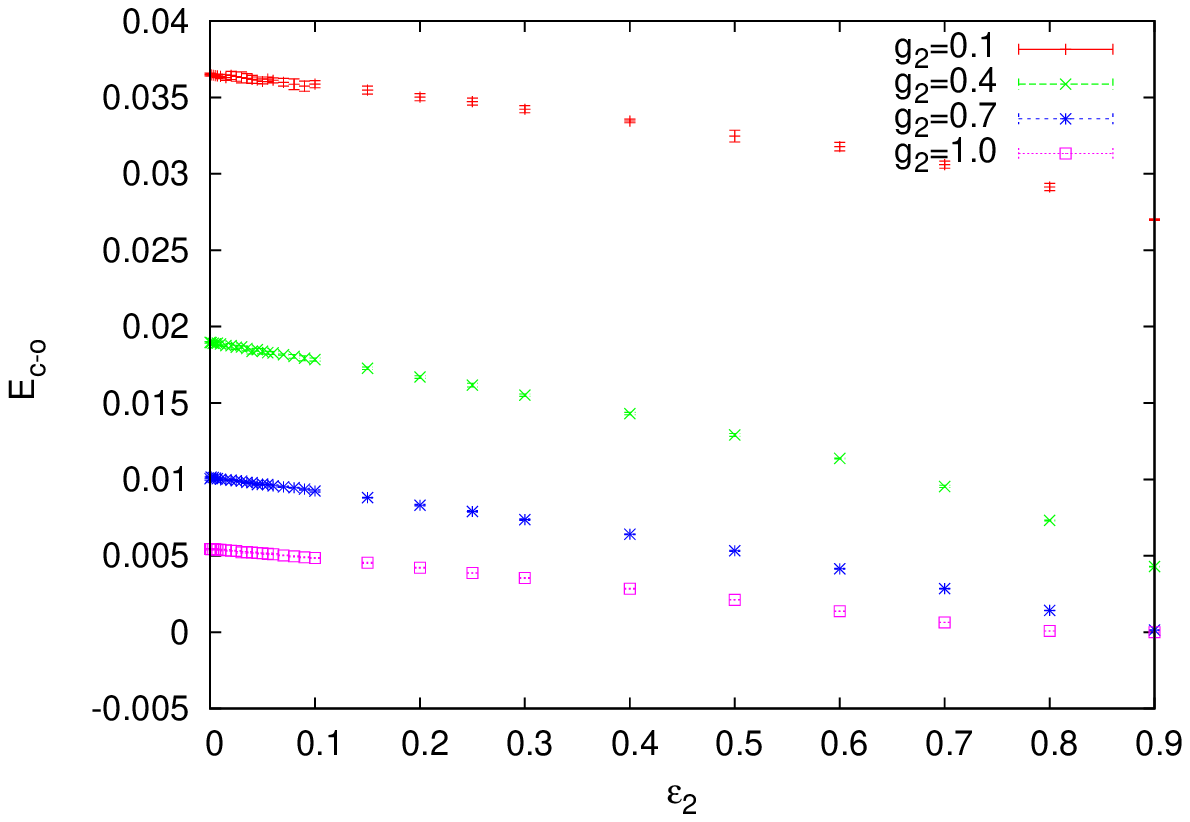}\label{eoc:1}} 
	\subfloat[]{\includegraphics[width=8cm]{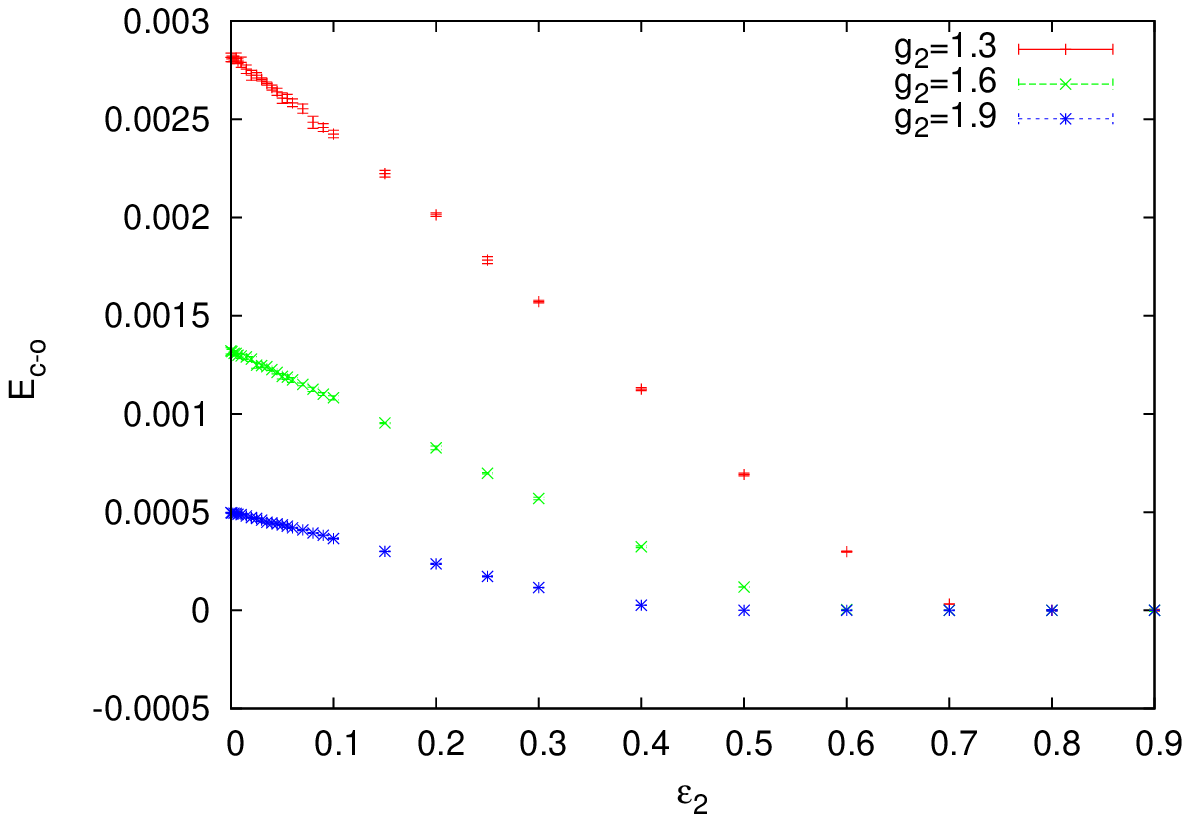}\label{eoc:2}}
	\caption{We show the energy range of the stable magnetic region $E_{c-o}$ 
in dependence on $\varepsilon_2$ for several values of $g_2$ and $n_1=n_2=1$.}
\label{eoc}
\end{figure}

\begin{figure}[!h]
\centering
	\subfloat[]{\includegraphics[width=8cm]{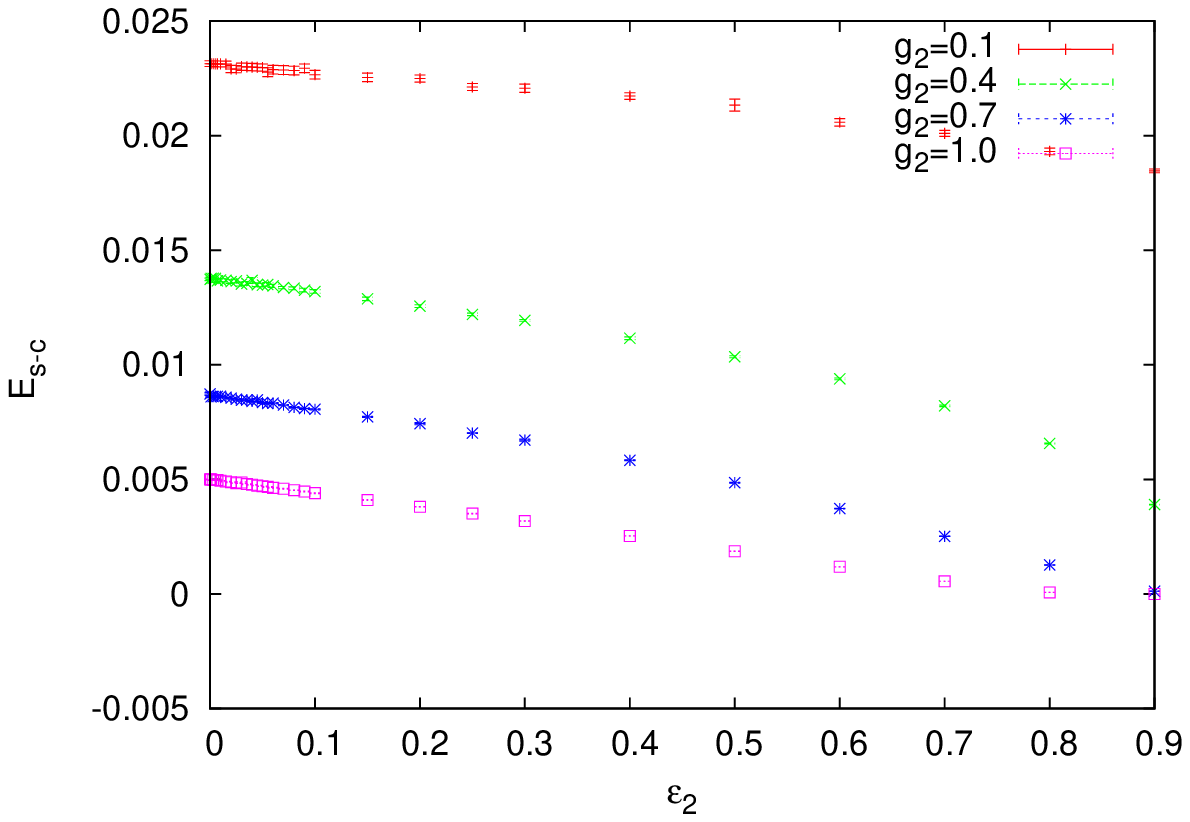}\label{esc:1}} 
	\subfloat[]{\includegraphics[width=8cm]{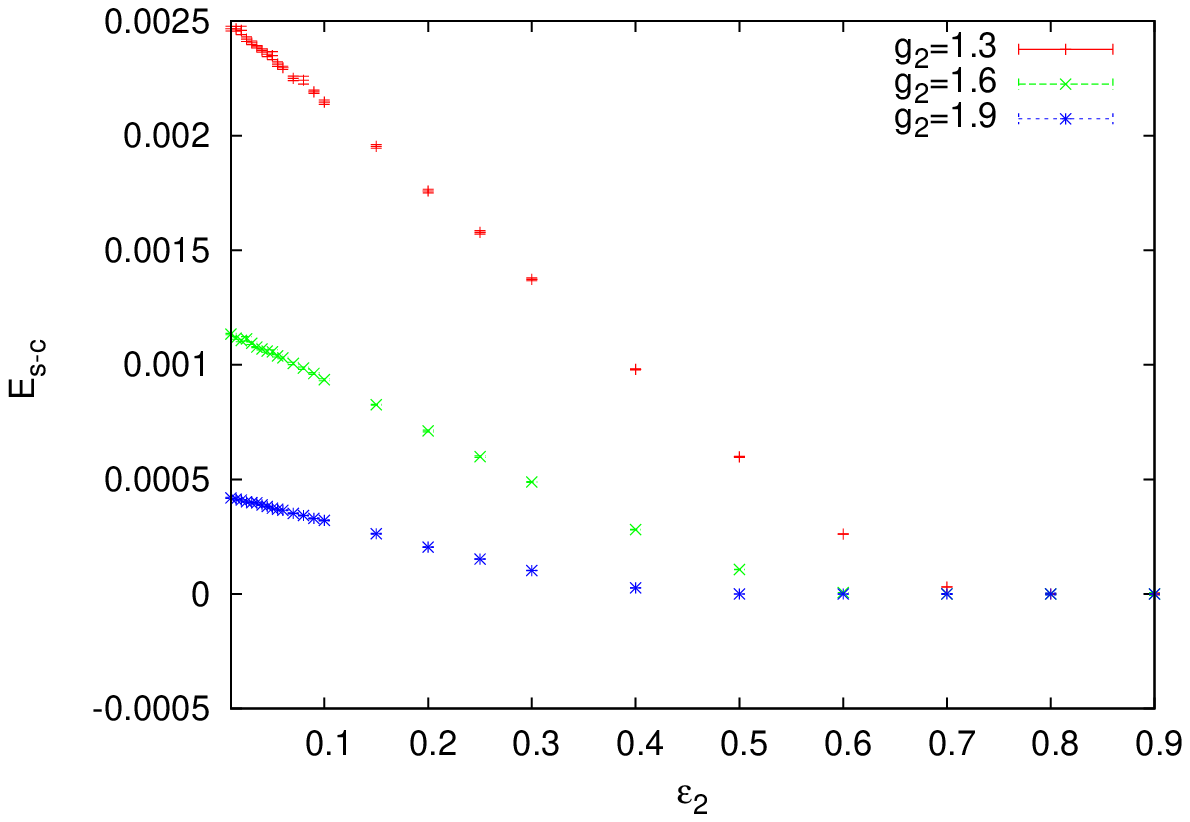}\label{esc:2}}
	\caption{We show the energy differences between current 
markers ``s'' and ``c'' given by $E_{s-c}$ in dependence on $\varepsilon_2$
for several values of $g_2$ and $n_1=n_2=1$.}
\label{esc}
\end{figure}

It is also interesting to note that for certain combinations of $(g_2, \varepsilon_2)$ 
the energy widths $E_{m-o}, E_{c-o}, E_{s-c}$ vanish, 
which implies that the magnetic region completely vanishes for these parameter choices. 
The larger $g_2$, i.e the smaller the width of the flux tube of the superconducting string
with respect to that of the Abelian-Higgs string, the smaller the value of 
$\varepsilon_2=(\varepsilon_2)_{\rm cr}$ at which
the magnetic regime disappears and no space-like currents are possible. We find e.g.
$(\varepsilon_2)_{\rm cr}\approx 0.73$ for 
$g_2=1.3$, $(\varepsilon_2)_{\rm cr}\approx 0.62$ for $g_2=1.6$
and $(\varepsilon_2)_{\rm cr}\approx 0.47$ for $g_2=1.9$.

Even though $E_{s-c}$ decreases with increasing $\varepsilon_2$ this is related to the overall 
shrinking of the entire magnetic branch. In order to understand how the position of
point ``s'' changes relative to the other current markers within the magnetic region, 
the tension-energy derivative (negative propagation velocity for longitudinal waves) 
at this point is plotted as a function of $\varepsilon_2$ in Fig.\ref{sDerivEps}. 

\begin{figure}[!h]
\centering
	\subfloat[]{\includegraphics[width=8cm]{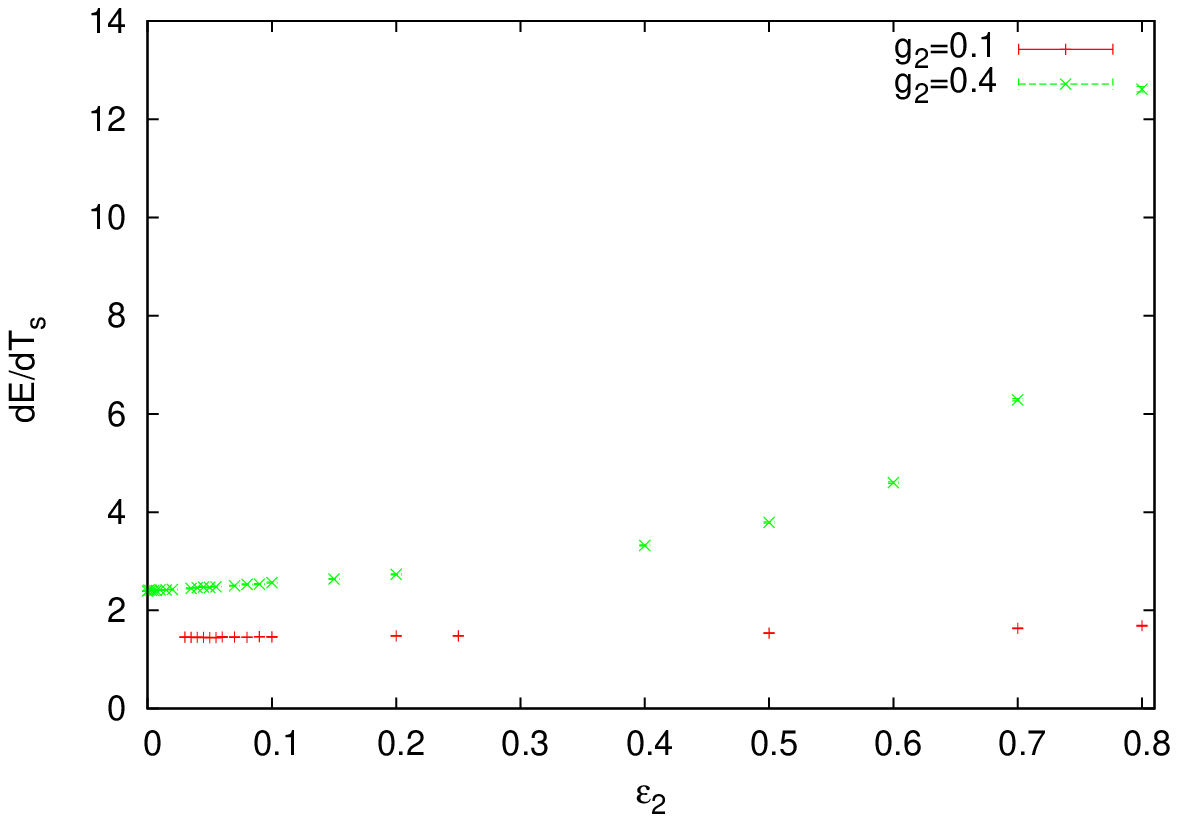}\label{sDerivEps:1}} 
	\subfloat[]{\includegraphics[width=8cm]{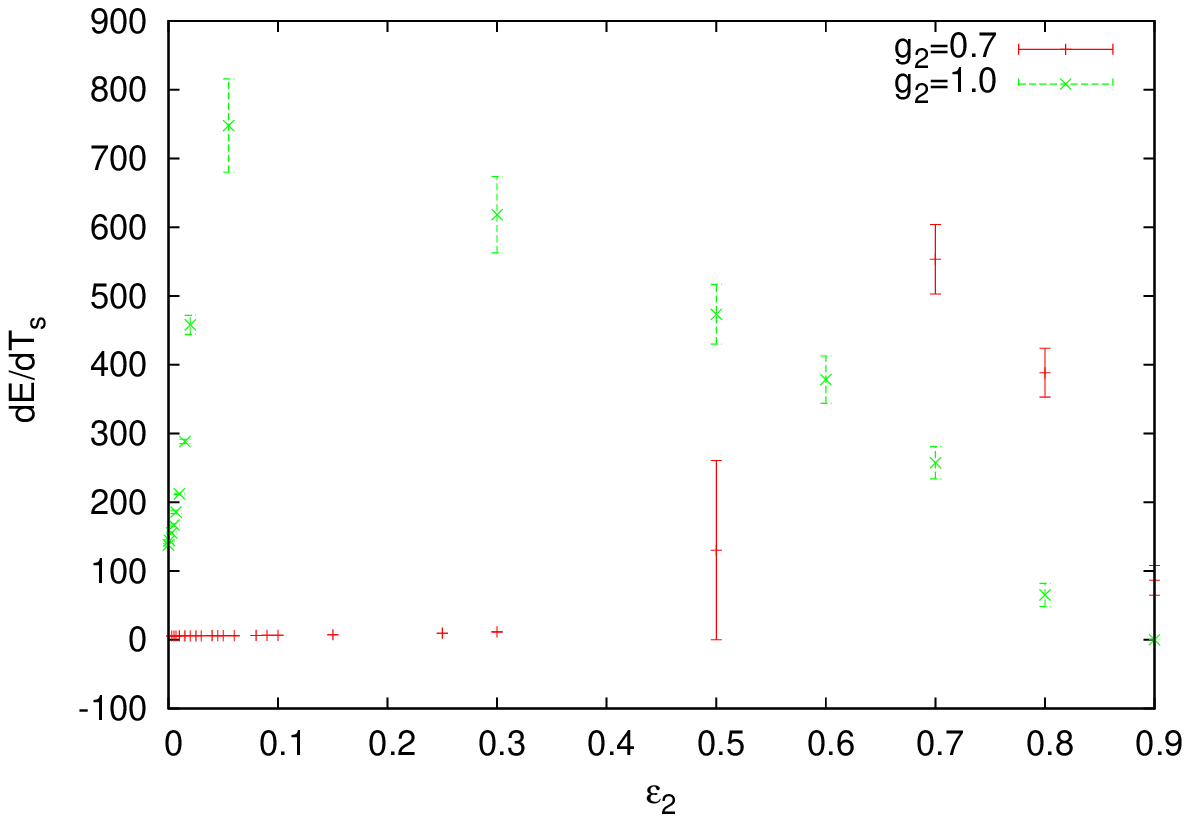}\label{sDerivEps:2}} \\
	\caption{The value of the derivative at current marker ``s'' 
$\left(\frac{dT}{dE}\right)_{s}$ in dependence of $\varepsilon_2$ 
for several values of $g_2$ and $n_1=n_2=1$. The bars indicate the errors 
in the numerical evaluation of the derivative.}
\label{sDerivEps}
\end{figure}

One can see that the derivative is always positive and increases with $\varepsilon_2$ until 
reaching the upper, magnetic limit where ``s'' coincides with ``m''. 
For $g_2>1.0$, this already happens at $\varepsilon_2 = 0$. 
Our results hence imply that -- at least for the potential parameters used here \eqref{PotParams} -- 
the condition $\eqref{VB}$ is falsified (if at all) only in the region where the string 
is already unstable according to criterion \eqref{stabil}.  


\subsubsection{Effect of winding numbers $n_1$ and $n_2$}
\label{I2-wind}

Here we fix $g_2=1$. In this case the energy $E$ and tension $T$ are simply given by $E = T = 2\pi(n_1 + n_2)$ 
when the current is absent or neutral, i.e. at points ``m'' and ``o'', respectively, 
if $\varepsilon_2=0$.
This is related to the fact that the choice $\beta_1=2.0$ and $\beta_2=2.0$ corresponds
to the Bogomolnyi-Prasad-Sommerfield (BPS) limit if the two strings do not carry currents
and are not coupled.

\begin{figure}[!h]
\centering
	\subfloat[$g_2=1.0$]{\includegraphics[width=8cm]{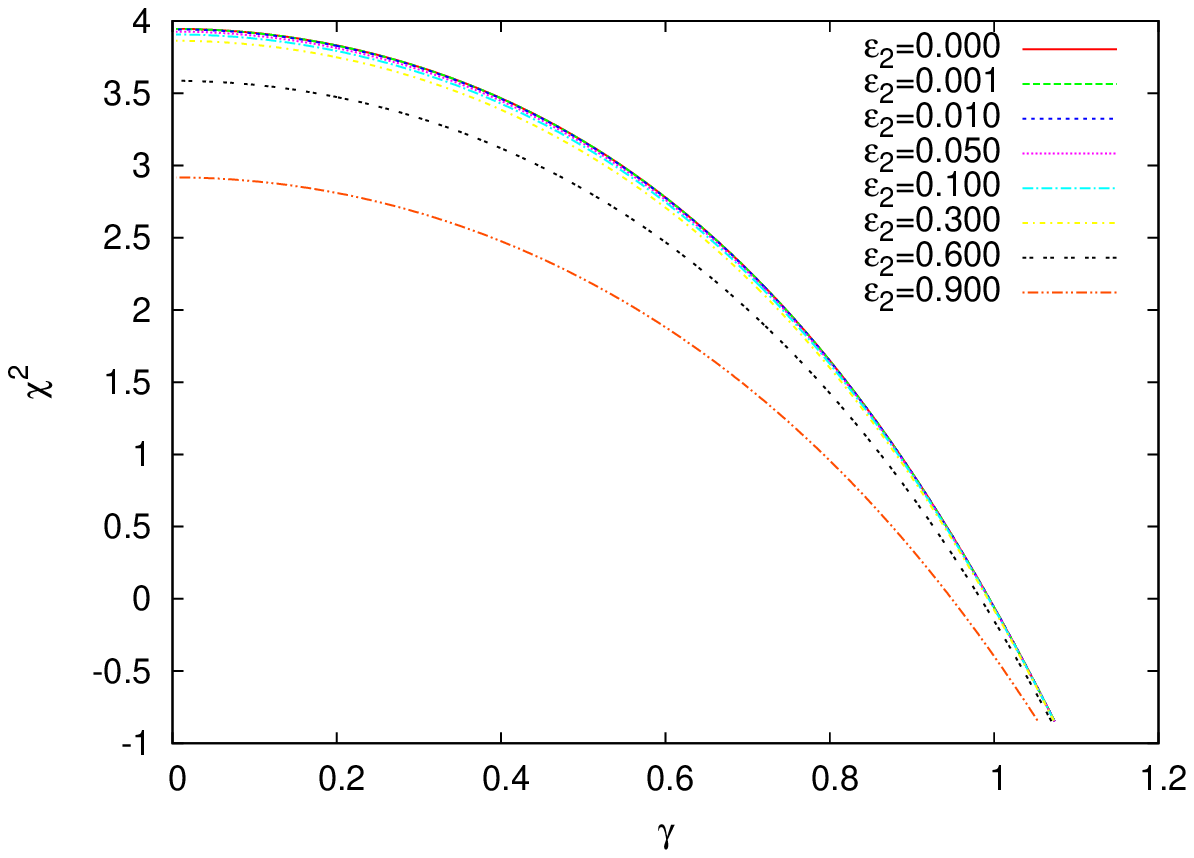}\label{first:1}} 
	\subfloat[$g_2=1.0$]{\includegraphics[width=8cm]{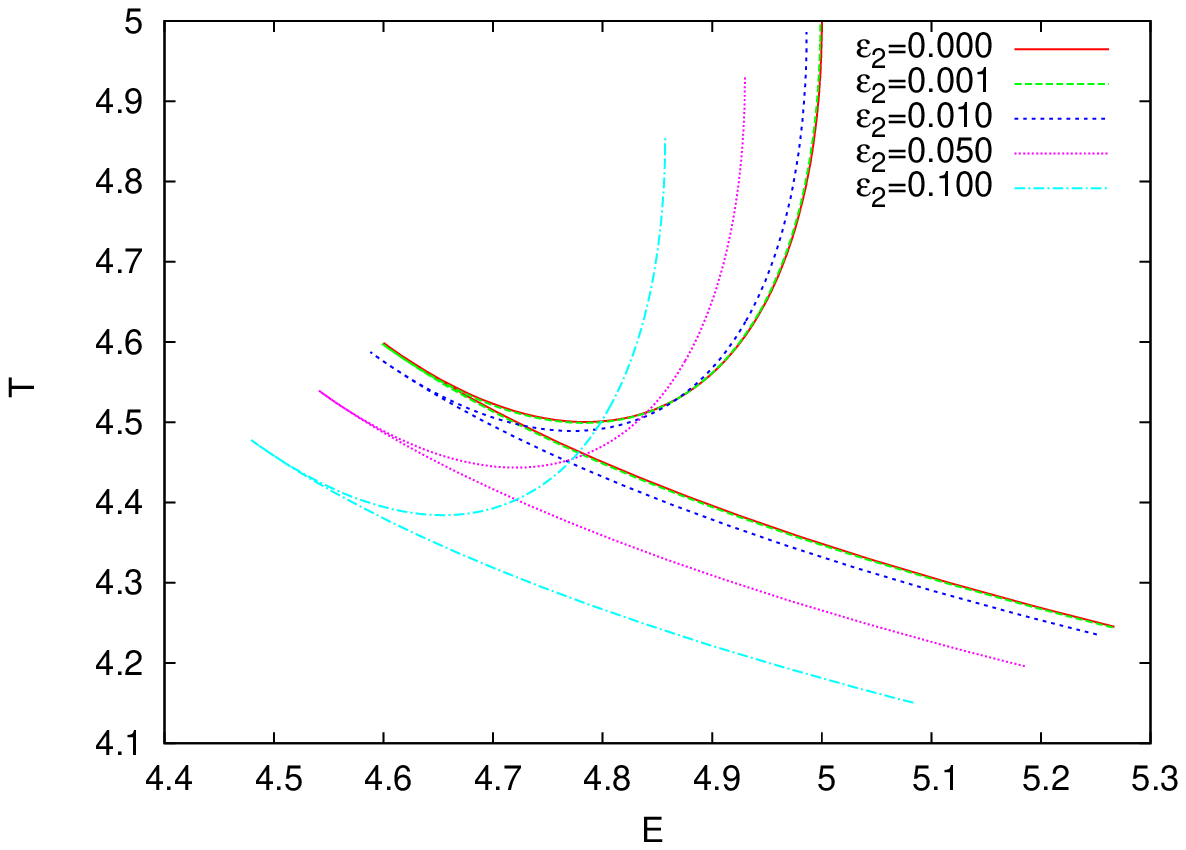}\label{first:2}}
	\caption{We show the influence of $\varepsilon_2$ 
on $\chi^2$ \subref{first:1} and on $T(E)$ \subref{first:2} for $n_1=2$ and $n_2=3$ and $g_2=1$. }
\label{first}
\end{figure}

First of all, consider the current nature and tension-energy diagrams for $n_1 = 2, n_2 = 3$ as shown in Fig. \ref{first}. 
When comparing this with Fig.\ref{ge} it becomes clear that 
increasing the windings produces qualitative and quantitative changes. We observe that 
an increase in the windings, which corresponds to an increase in the magnetic fluxes 
$\Phi_1$, $\Phi_2$,
leads to an increase in the maximal possible current on the superconducting string
before the current decouples ($\gamma=0$ and hence $f_3(x)\equiv 0$). Moreover, 
the value of $\gamma$ at the phase frequency threshold $\chi^2_e\approx -0.848$ 
increases.    
In order to understand the behaviour of the current regions for different combinations of the windings, 
we have plotted $E_{e-o}(\varepsilon_2)$, $E_{m-o}(\varepsilon_2)$  
and $E_{c-o}(\varepsilon_2)$ in Fig.\ref{eon}, Fig.\ref{mon} and Fig.\ref{con}, respectively, 
for all combinations of  $(n_1,n_2) \in \{1,2,3,4\} \times \{1,2,3,4\}$ and $g_2=1$. 

\begin{figure}[!h]
\centering
	\subfloat[$n_2 = 1$]{\includegraphics[width=8cm]{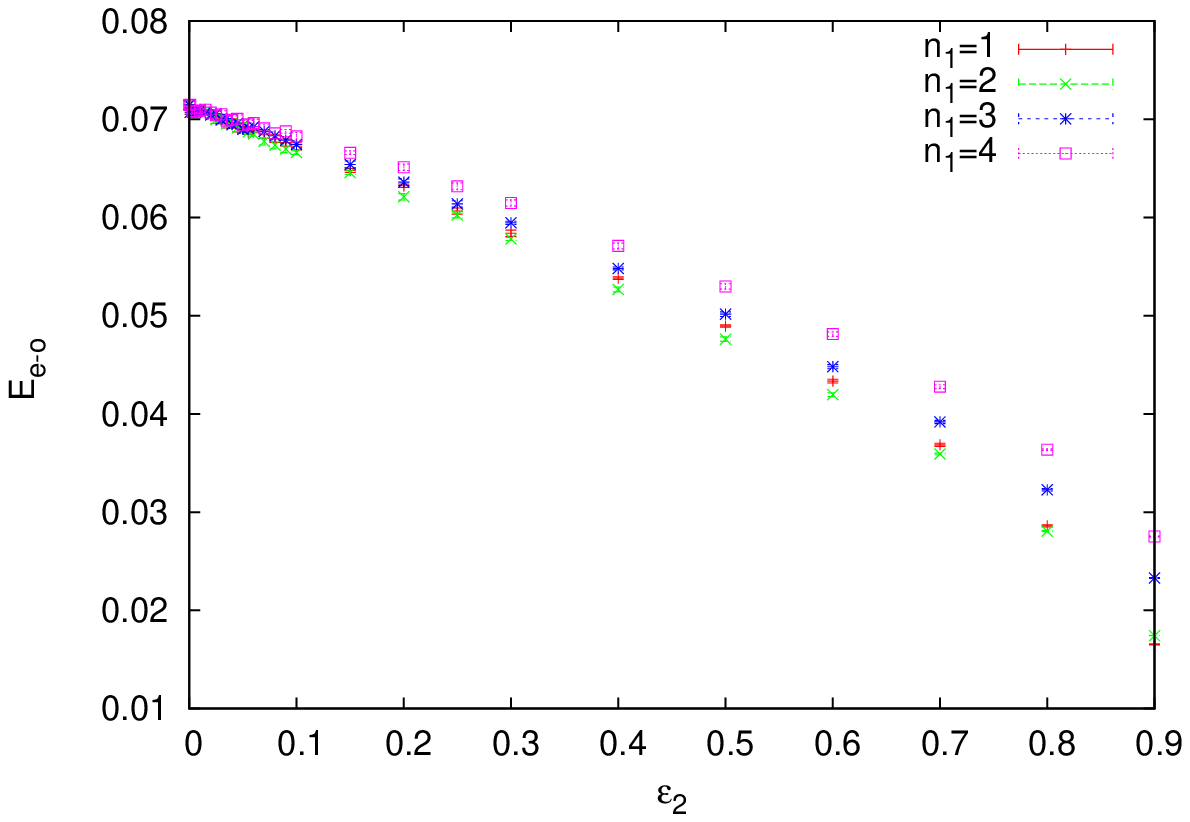}\label{eon:1}} 
	\subfloat[$n_2 = 2$]{\includegraphics[width=8cm]{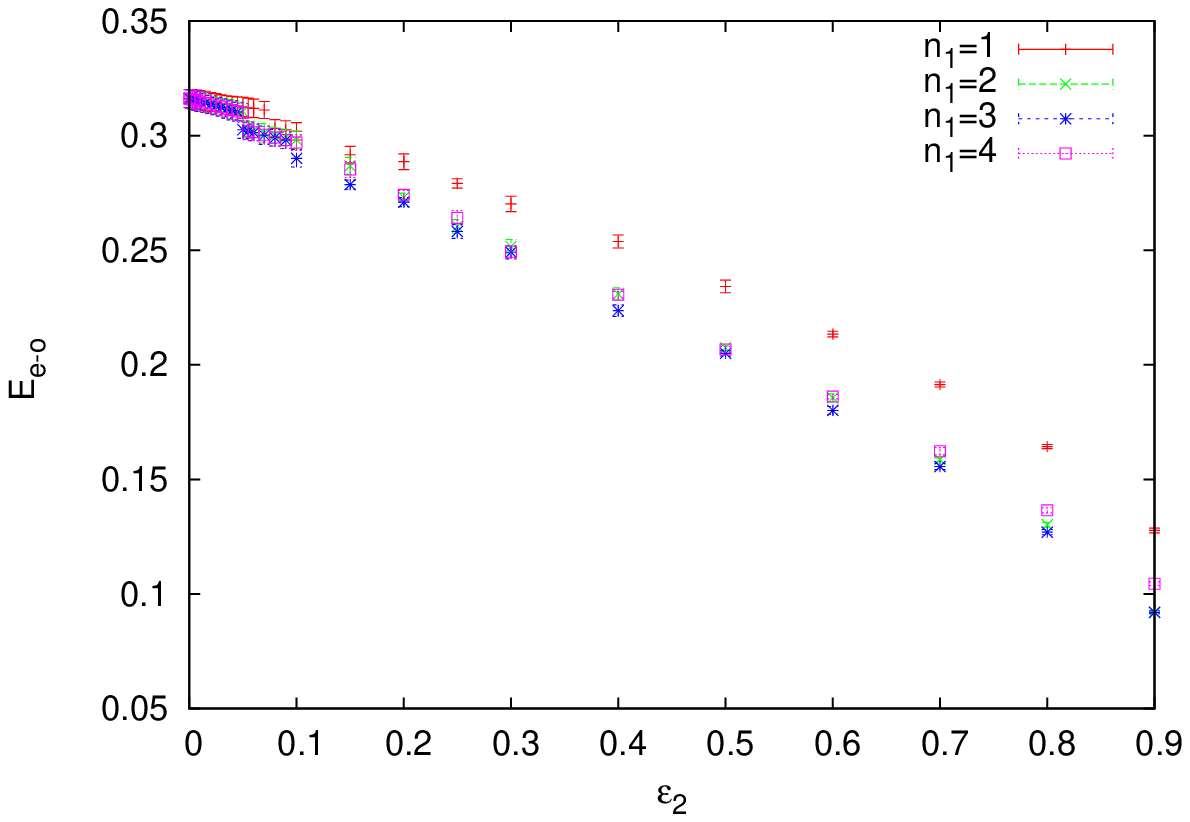}\label{eon:2}}\\
	\subfloat[$n_2 = 3$]{\includegraphics[width=8cm]{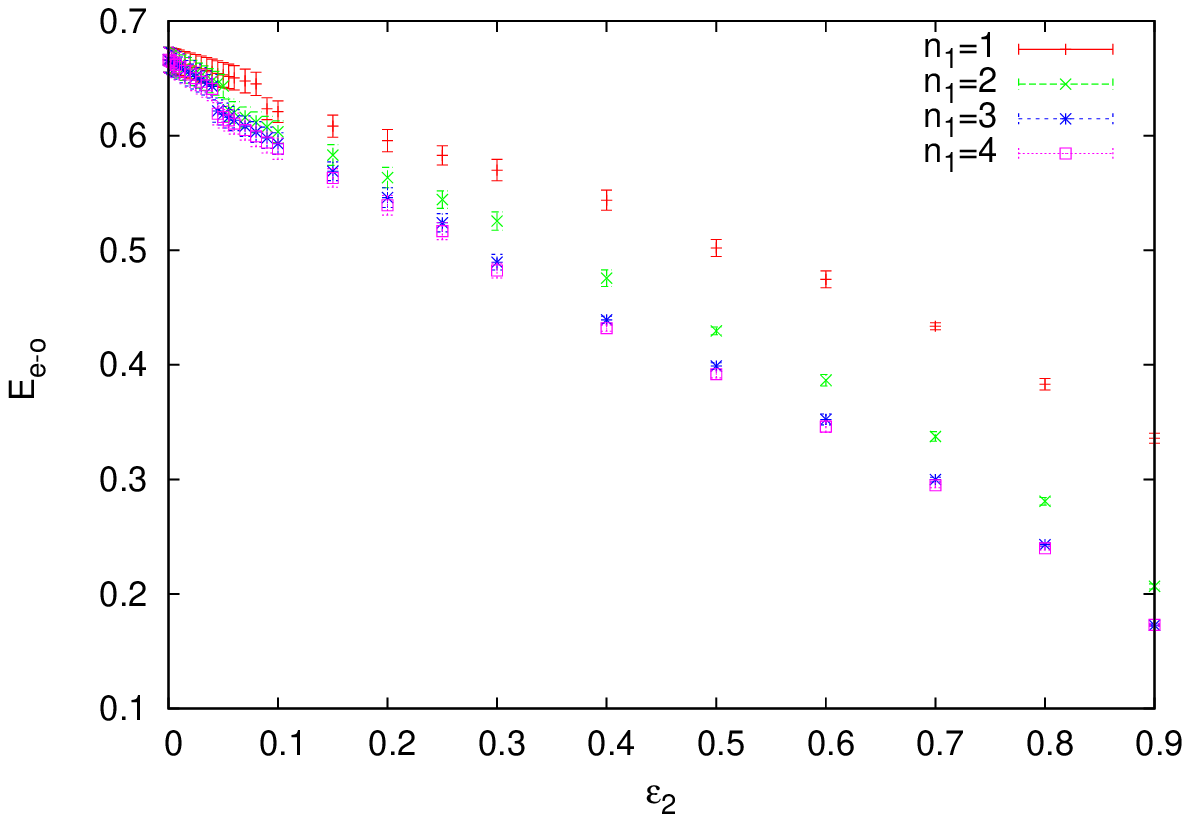}\label{eon:3}}
	\subfloat[$n_2 = 4$]{\includegraphics[width=8cm]{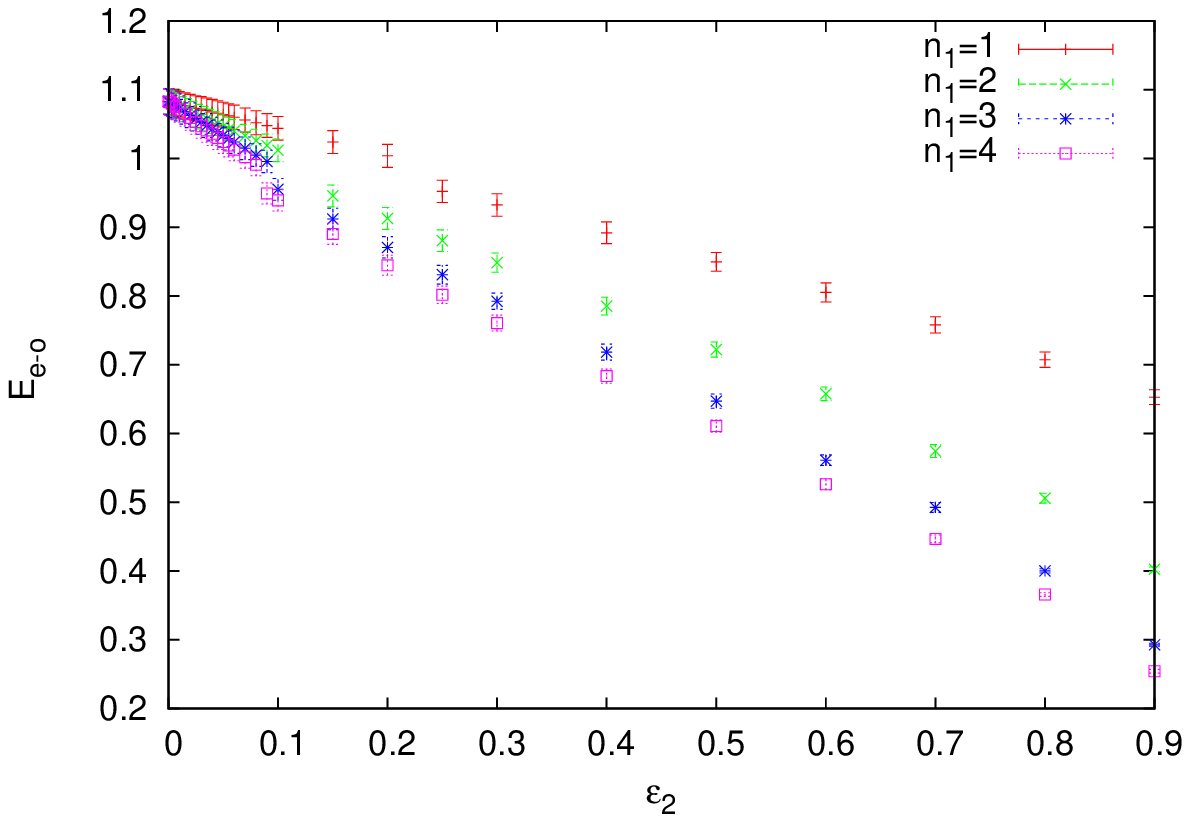}\label{eon:4}}
	\caption{We show the energy range of the electric region $E_{e-o}$ 
in dependence on $\varepsilon_2$ for $n_1, n_2=1,2,3,4$ and $g_2=1$.}
\label{eon}
\end{figure}

\begin{figure}[!h]
\centering
	\subfloat[$n_2 = 1$]{\includegraphics[width=8cm]{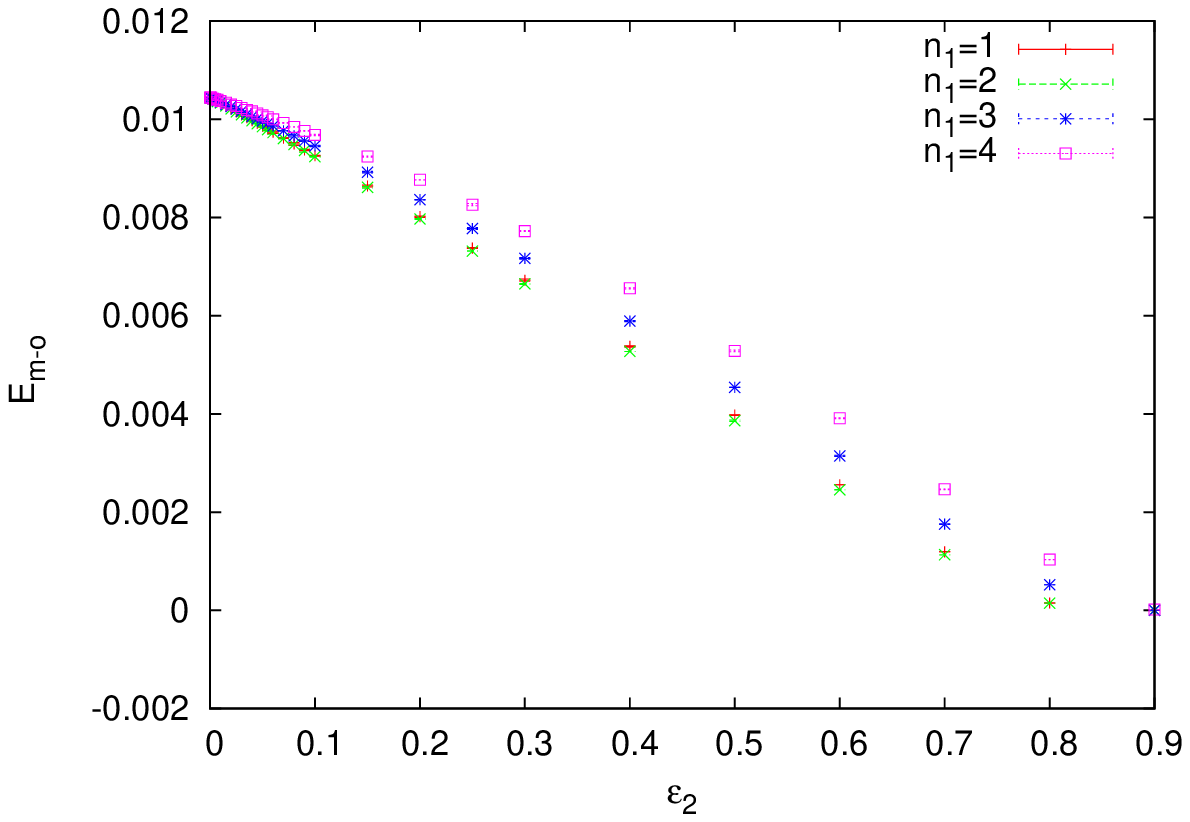}\label{mon:1}} 
	\subfloat[$n_2 = 2$]{\includegraphics[width=8cm]{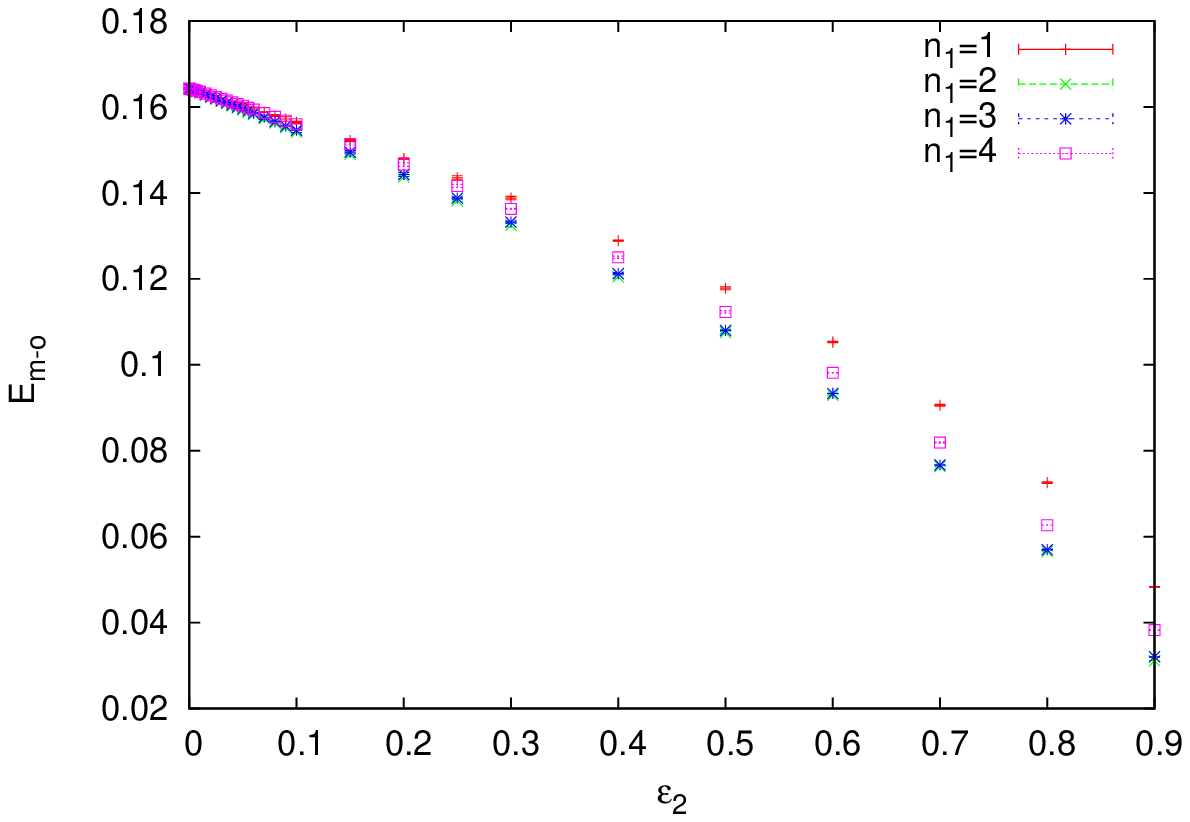}\label{mon:2}}\\
	\subfloat[$n_2 = 3$]{\includegraphics[width=8cm]{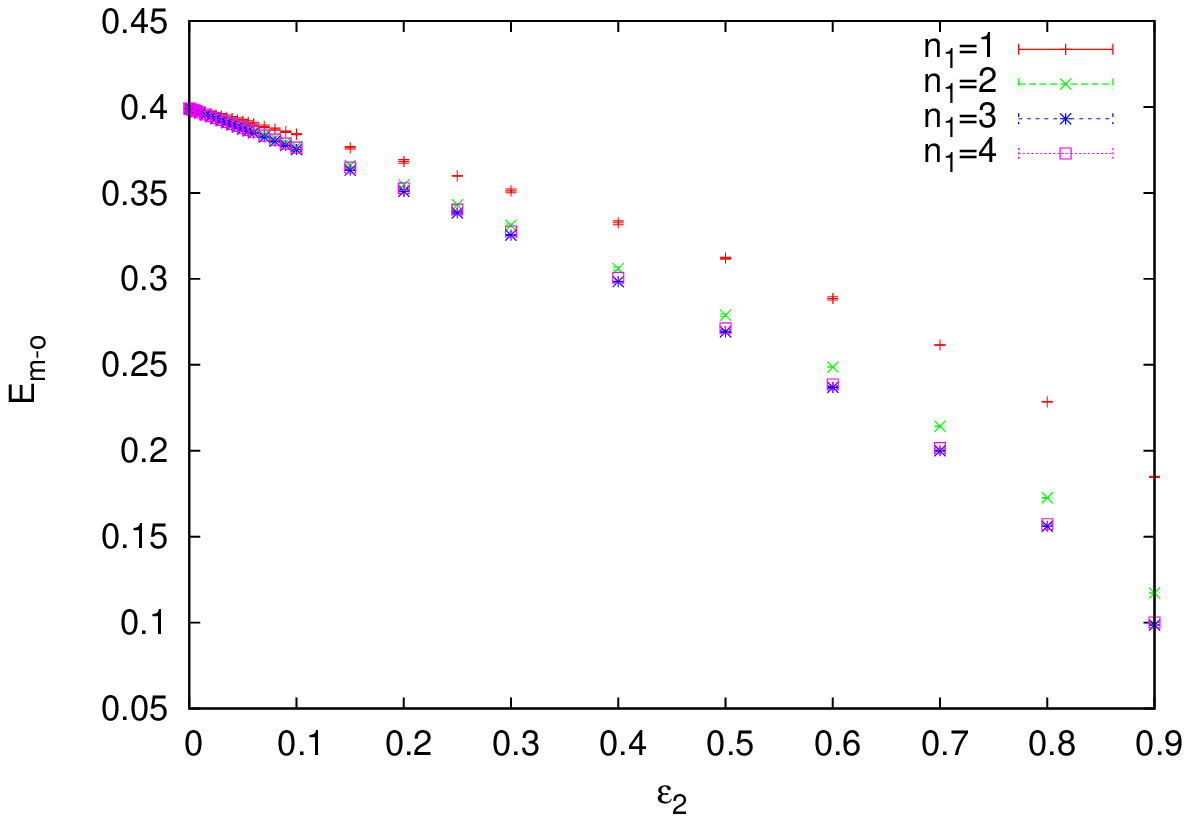}\label{mon:3}}
	\subfloat[$n_2 = 4$]{\includegraphics[width=8cm]{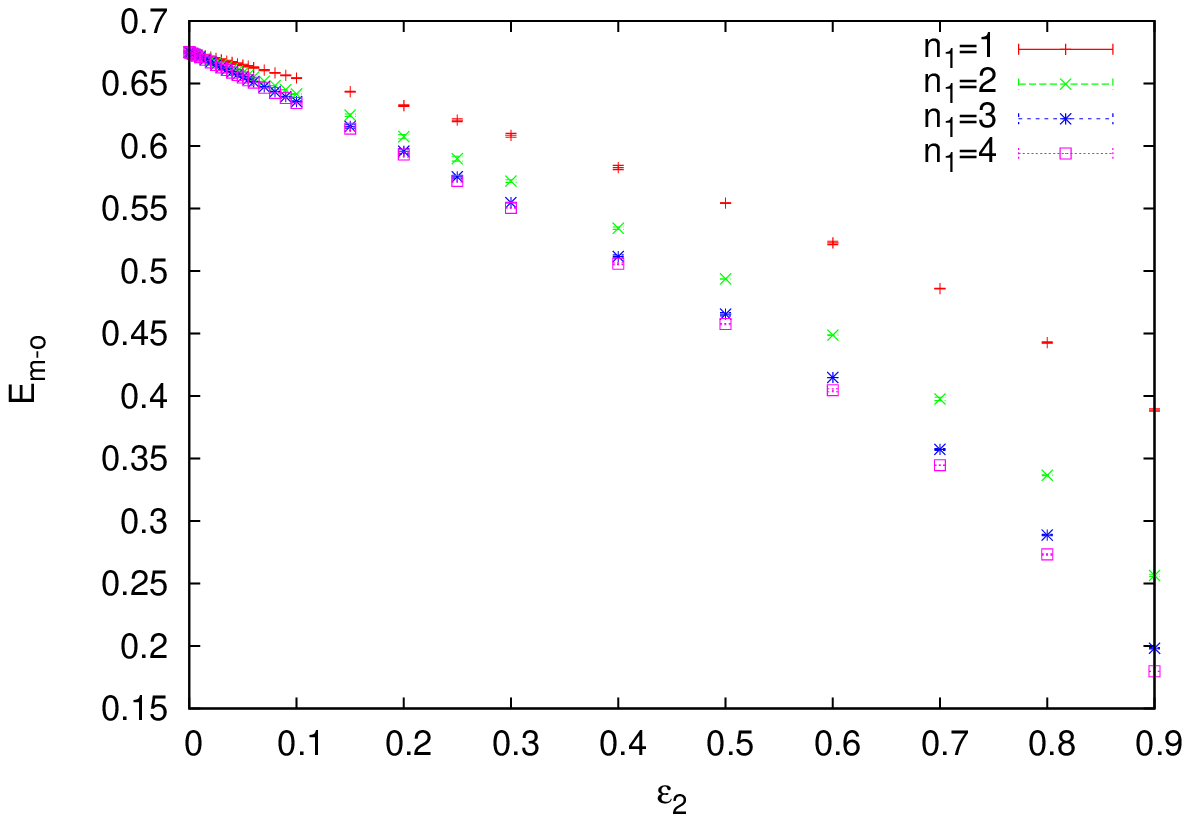}\label{mon:4}}
	\caption{We show the energy range of the 
magnetic region $E_{m-o}$ in dependence on $\varepsilon_2$ for $n_1, n_2=1,2,3,4$ and $g_2=1$.}
\label{mon}
\end{figure}
 
\begin{figure}[!h]
\centering
	\subfloat[$n_2 = 1$]{\includegraphics[width=8cm]{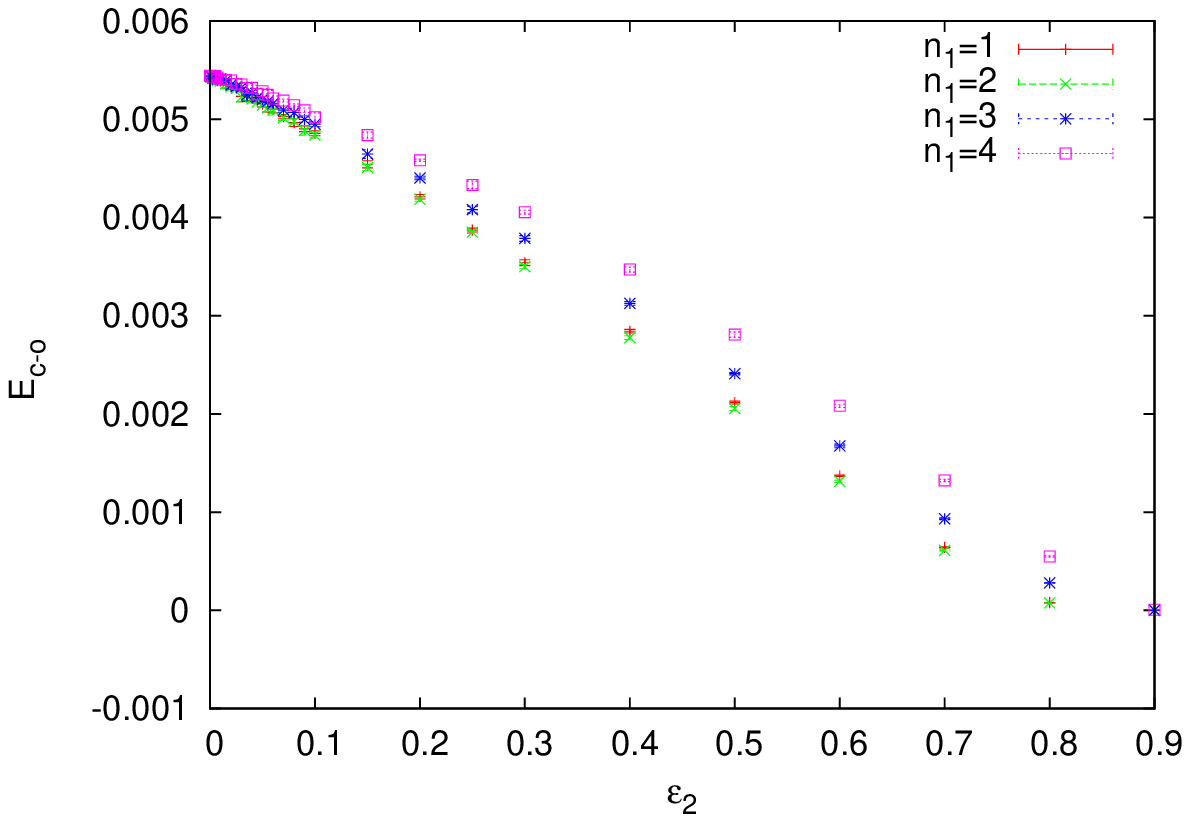}\label{con:1}} 
	\subfloat[$n_2 = 2$]{\includegraphics[width=8cm]{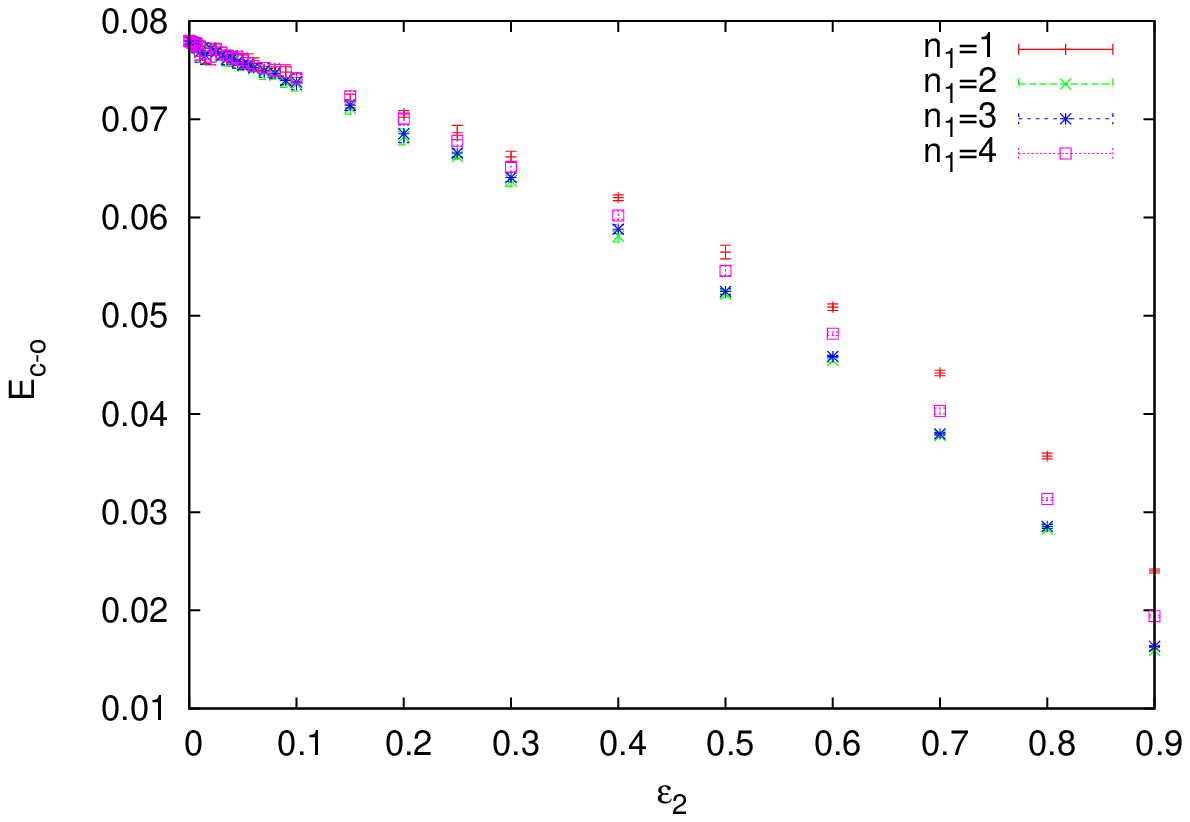}\label{con:2}}\\
	\subfloat[$n_2 = 3$]{\includegraphics[width=8cm]{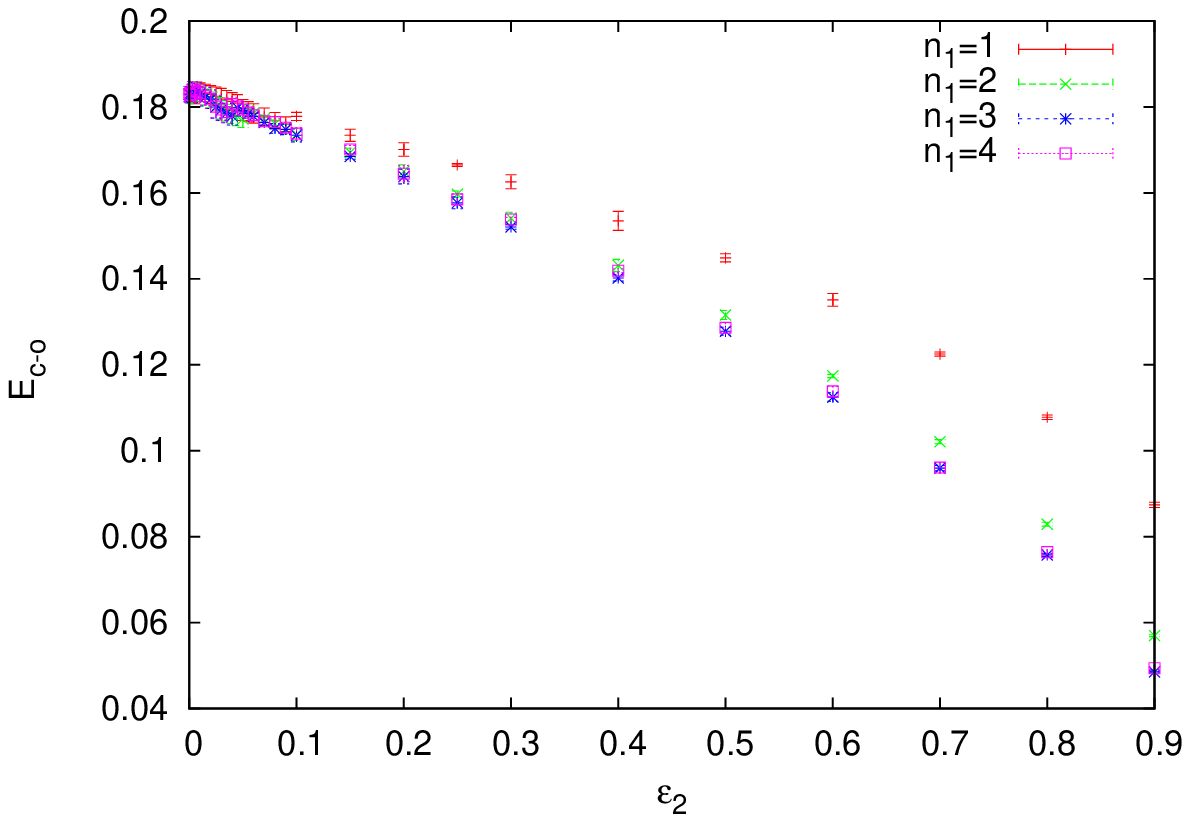}\label{con:3}}
	\subfloat[$n_2 = 4$]{\includegraphics[width=8cm]{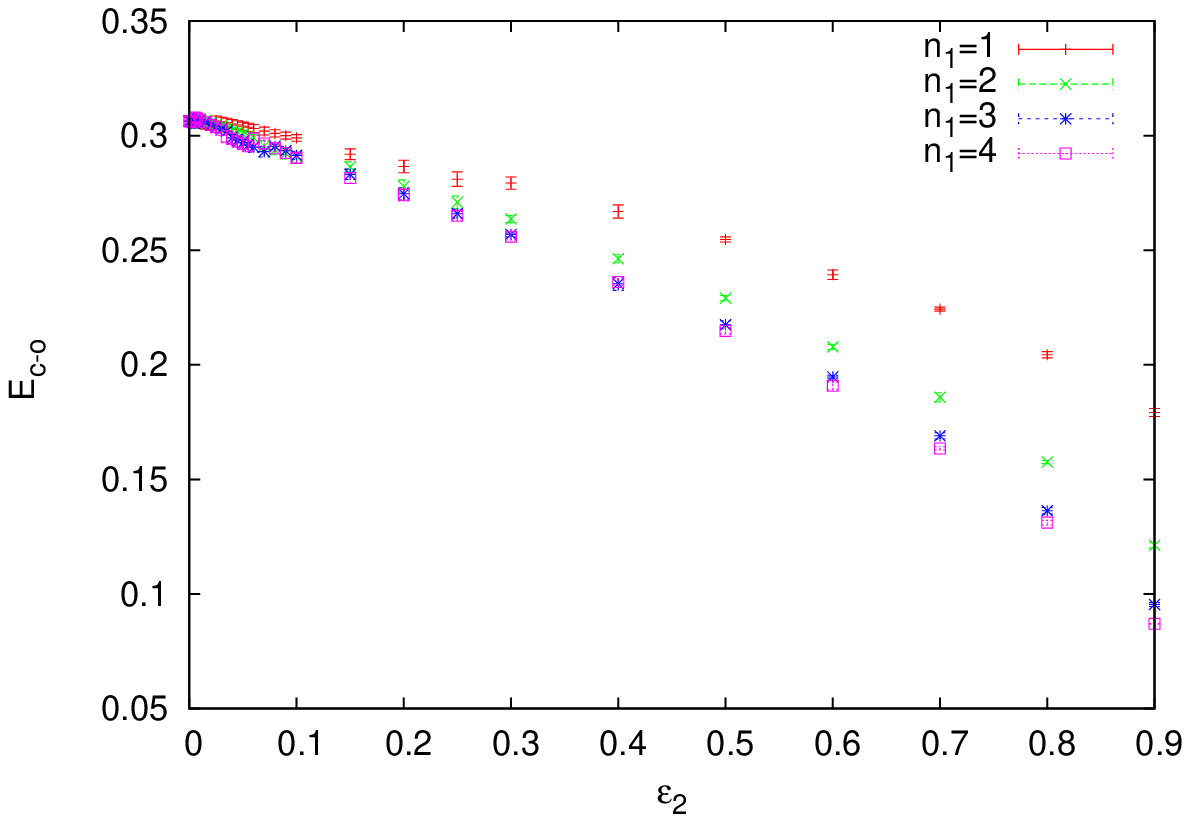}\label{con:4}}
	\caption{We show the energy range of the stable magnetic region $E_{c-o}$ 
in dependence on $\varepsilon_2$ for $n_1,n_2=1,2,3,4$ and $g_2=1$.}
\label{con}
\end{figure}

We observe that the overall energy widths of all current regions dramatically increase with 
the increase of $n_2$. Moreover, an increase in $n_1$ produces a shift of all curves, 
which also increases with $\varepsilon_2$. 
However, the ordering of the curves for a certain $n_2$ is not solely dependent on 
$n_1$, but also on $n_2$, even though we could not deduce the exact ordering rules (if any) 
from our numerical data. 
Nonetheless, these effects should be connected with the observation made 
in \cite{BHFA} stating that strings of equal windings bind stronger. From this one
is led to conclude that for a given $n_2$ the energy widths are smallest for $n_1=n_2$.
However, we only observe this to be strictly true for $n_2=4$.

\section{Conclusions and Outlook}
\label{conclusions}
In this paper we have studied the interaction of superconducting strings with 
Abelian-Higgs strings via an attractive gauge field interaction and have investigated the stability
of the superconducting strings. We have chosen the unbroken U(1) symmetry of the superconducting
string to be ungauged in order to maintain the localized character of the solution in accordance
with \cite{PP}. We find that the change of the interaction parameter has similar effects
than the change of the ratio between the radius of the flux tube of the superconducting
and that of the Abelian-Higgs string. Increasing either the interaction parameter
and hence increasing the attractive interaction between the superconducting
and Abelian-Higgs string or the ratio of the widths leads to a decrease
in the maximal possible current on the superconducting string and a decrease
of the condensate on the string axis given by $f_3(0)\equiv \gamma$ 
at the phase frequency threshold beyond which no stationary solutions exist and at which
the tension and energy diverge. While the strings are stable with respect to
longitudinal and transverse perturbation if they carry time-like currents, there is only a
limited domain in the case of space-like currents where they are stable 
with respect to longitudinal
perturbations. We find that space-like currents are not possible for sufficiently strong interaction between the superconducting
and Abelian-Higgs string and/or small enough ratio between the widths of the flux tubes.
Increasing the windings and hence the magnetic fluxes on the strings leads to an
increase in the maximal possible current on the string and an increase 
of the condensate on the string axis $f_3(0)=\gamma$ at the phase frequency threshold.

Our results have important implications for vorton formation in models where cosmic strings
are coupled to superconducting strings. In the case of (p,q)-strings a field theoretical
model has been developed that describes the bound state of p F-strings and q D-strings
by two coupled Abelian-Higgs strings identifying p and q with the windings of the respective
strings \cite{saffin}. The model studied in this paper is an extension of this model
by allowing the D-string to carry a bosonic current which was shown to be possible
in special cases \cite{polchinski}.  
\\
\\
{\bf Acknowledgments}
We are grateful to P. Peter for discussions and comments on this manuscript.

\end{document}